\documentclass[aps,onecolumn,preprint,superscriptaddress]{revtex4}
\usepackage{graphics}
\usepackage{graphicx}
\usepackage{amssymb}		
\usepackage{amsmath}
\usepackage{verbatim}
\usepackage{axodraw}
\usepackage{pstricks}

\newcommand{\beq}{\begin{equation}}
\newcommand{\eeq}{\end{equation}}
\newcommand{\bea}{\begin{eqnarray}}
\newcommand{\eea}{\end{eqnarray}}

\def \et {E_{T}}
\def  \met {\not\!\!\et }

\begin{document}

\begin{flushright}MCTP-09-02 \\
\end{flushright}
\vspace{0.2cm}

\title{Cosmic Ray Positrons from Annihilations into a New, Heavy Lepton}

\author{Daniel J. Phalen}
\affiliation{Michigan Center for Theoretical Physics (MCTP) \\
Department of Physics, University of Michigan, Ann Arbor, MI
48109 
}
\author{Aaron Pierce}
\affiliation{Michigan Center for Theoretical Physics (MCTP) \\
Department of Physics, University of Michigan, Ann Arbor, MI
48109 
}
\author{Neal Weiner}
\affiliation{Center for Cosmology and Particle Physics\\
Department of Physics, New York University, New York, NY
10003
}

\date{\today}

\begin{abstract}
Recent results from the PAMELA experiment indicate an excess in the positron spectrum above 10 GeV, but anti-proton data are consistent with the expected astrophysical backgrounds.  We propose a scenario that reproduces these features. Dark matter annihilates through channels involving a new heavy vectorlike lepton which then decays by mixing with Standard Model leptons.  If charged, this heavy lepton might be produced at the LHC, and could lead to  multi-lepton final states or to long-lived charged tracks.  Large neutrino detectors such as ANTARES or IceCube might be sensitive to a monochromatic neutrino line. This scenario may be simply embedded in various models, including an extension to the NMSSM.  
\end{abstract}

\maketitle

\setcounter{equation}{0}

\section{Introduction}
 
There is great deal of excitement surrounding recent measurements of the positron flux by the PAMELA experiment \cite{Adriani:2008zr, Adriani:2008zq}, which show an increase in the positron fraction above $\sim$ 10 GeV.  Interactions of cosmic ray protons with the interstellar medium would be unable to produce a significant rise, prompting a frenzy of activity in the dark matter community as well as a reexamination of pulsars \cite{Hooper:2008kg,Yuksel:2008rf,Profumo:2008ms}, in light of earlier predictions \cite{aharonian:1995}.

Interpreted as a dark matter signal, the PAMELA result places strong constraints on annihilation channels. The sharp rise indicates a hard lepton spectrum \cite{Cirelli:2008pk, Cholis:2008hb}, while the lack of an accompanying rise in the $\bar{p}/p$ ratio strongly constrains hadronic annihilation modes \cite{Donato:2008jk}.  Consequently, if PAMELA results are taken at face value, the  dark matter particle annihilates primarily to leptons --- hadronic channels are suppressed.   

Construction of a dark matter candidate that primarily annihilates directly to leptons is non-trivial.  Majorana fermions have helicity suppressed annihilations to light fermions, and Dirac fermions face significant model building challenges \cite{Harnik:2008uu}. However, one can consider new annihilation modes into non-Standard Model states, which then subsequently decay to leptons. Such models can be divided into two categories.  In the first approach,  dark matter annihilates to light intermediate particles.  If sufficiently light ($\sim$ GeV), the absence of anti-protons  can be explained by simple kinematics \cite{Cholis:2008vb} (see also \cite{ArkaniHamed:2008qn, Pospelov:2008jd,Nelson:2008hj, Cholis:2008qq,Nomura:2008ru}). A second approach is for the dark matter to preferentially couple to leptons via dynamics \cite{Fox:2008kb}.  Alternative models have considered the possibility where dark matter {decays preferentially to leptons \cite{Chen:2008md,Chen:2008qs,Hamaguchi:2008ta,Ibarra:2008jk,Arvanitaki:2008hq,Chen:2008yi,Ishiwata:2008cv,Yin:2008bs,Zhang:2008tb}.

Here we present a hybrid approach.  Dark matter dominantly annihilates to final states involving new, heavy particles.  These particles mix with the Standard Model leptons, and so carry lepton number.  Following the decay of these new particles, dark matter annihilations produce hard leptons in association with gauge bosons.  The annihilation spectra from this model are distinct from those previously included in the literature thus far. 

In the next section we discuss the general framework for dark matter annihilations and heavy lepton decays.  The scenario is a very simple extension of the Standard Model. In section III, we calculate the cosmic ray fluxes from dark matter annihilations and the boost factor relative to the thermal cross section.  We find this scenario can explain the PAMELA excess with a small boost factor.  In section IV, we discuss the possible observation of high energy neutrinos from dark matter annihilations.  In section V, we discuss the possible production of new vector leptons at the LHC, and their signatures.  In section VI we construct extensions of the Standard Model and Supersymmetric Standard Model which realize this scenario.  Finally, we conclude.

\section{Annihilations of dark matter into a New, Heavy Lepton \label{sec:annihilations}}
The PAMELA data do more than confirm the excess seen by HEAT \cite{Barwick:1997ig} and AMS-01 \cite{Aguilar:2007yf}. The sharp upturn in the positron fraction seems to disfavor annihilations with hadronic modes, because the positrons from a hadronic shower are generally too soft to produce such a spectrum.  Furthermore,  such showers contain anti-protons, and no excess is observed in this channel. Thus, to reproduce the positron/electron ratio and the antiproton/proton ratio observed by PAMELA, dark matter annihilations should produce a larger ratio of high energy positrons to antiprotons than annihilations to vector bosons \footnote{See, however, \cite{KanePiercePhalen}  for a discussion of whether uncertainties in propagation parameters might accommodate  annihilation to gauge bosons.}.

We propose a new mechanism to achieve this. 
We assume dark matter annihilation proceeds via the processes in Figure \ref{fig:DMannihilation}.  The dark matter particle, $\chi$, annihilates through new channels which involve heavy vectorlike states, $\Xi / \bar\Xi$,  which carry lepton number and thus subsequently decay into light leptons and $W$ or $Z$ bosons.

\begin{figure}
  \begin{center}
    \begin{picture}(400,200)(0,0)
      \ArrowLine(50,50)(100,100) 
      \Text(45,45)[rt]{$\chi$}
      \ArrowLine(100,100)(50,150) 
      \Text(45,155)[rb]{$\chi$}
      \ArrowLine(150,50)(100,100) 
      \Text(155,45)[lt]{$\bar{\Xi}$}
      \ArrowLine(100,100)(150,150) 
      \Text(155,155)[lb]{$\Xi, l$}
      \GCirc(100,100){20}{0.5}   
      \ArrowLine(250,100)(300,100) 
      \Text(245,100)[r]{$\Xi$}
      \ArrowLine(300,100)(350,50) 
      \Text(355,45)[lt]{$\mathit{l}$}
      \Photon(300,100)(350,150){5}{5} 
      \Text(355,155)[lb]{$W,Z$}
    \end{picture}
  \end{center}
  \caption{{\it Left:} The annihilation process of the dark matter $\chi$ to a heavy $\bar \Xi$ lepton and either $\Xi$ or a Standard Model lepton $l$.  The shaded circle represents a model-dependent annihilation mechanism.  We discuss possibilities for this in section \ref{sec:Models}.
{\it Right:} Decay of the $\Xi$ into to a lepton and a $W$ or $Z$ boson. \label{fig:DMannihilation}}
\end{figure}
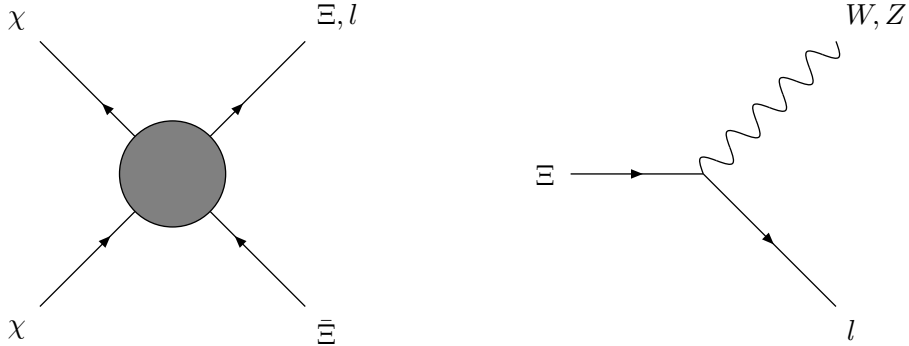

This theory can explain the PAMELA signal. The introduction of new, heavy leptons allows direct annihilation into leptonic states which are not suppressed by a small fermion mass. This can occur either from direct annihilations to heavy leptons alone (i.e. through a $\chi \chi \bar \Xi \Xi$-type operator) or from a combination of new heavy leptons and light leptons (through a $\chi \chi \bar \Xi l_i$-type operator). Because we do not rely upon a helicity flip on the external line, such processes can be large and dominate the annihilation rate. We will not specify the mechanism by which the annihilation proceeds for now, but will revisit this question.  (A similar annihilation to a heavy/light pair can occur in RS theories, although with different masses and spectra than we consider here \cite{Ponton:2008zv}.)

Under $SU(2) \times U(1)$ the $\Xi$ can have quantum numbers $({\bf 2 }\, , {\bf  1/2})$ (like a Standard Model $\ell$), $({\bf 1 }\, , {\bf  1})$ (like a Standard Model $e^c$), or $({\bf 1 }\, , {\bf  0})$ (a complete Standard Model singlet, $n$). Any of these is viable, but they have distinct phenomenology.

In the case where we add vectorlike particles with quantum numbers of $e^c$ or $n$, decays proceed with comparable rates both through $W$ and $Z$ bosons\footnote{We thank Tom Rizzo for a discussion of this point.}. Thus, for $\chi \chi \rightarrow \Xi \bar{\Xi}$, each annihilation (on average) roughly yields a single charged lepton and two gauge bosons. In contrast,  as we shall see, annihilation to a pair of vectorlike heavy doublets ultimately yields final states with charged leptons and gauge bosons in a one to one ratio.  Since gauge bosons are a source of anti-protons, taking $\Xi$ as a vectorlike $SU(2)_L$ doublet produces the highest $e^+/{\bar p}$ signal, as is desirable to fit the PAMELA data.  

For the study of cosmic ray signals, then, we will focus on the $\ell$-like case. For the purpose of the PAMELA data, the consequence of considering the $n$- or $e ^c$-like models instead is to multiply the $\bar p$ spectrum of the case at hand by a factor of $\sim 2$, as an increased boost factor would be needed to fit the positron data (see section~\ref{sec:pamela}).

For concreteness, we begin with a Lagrangian\footnote{A similar Lagrangian can be constructed for the situation where $\Xi$ is neutral (see, for instance, \cite{spencer}). }
\beq
\mathcal{L} = \sum_i y_i \ell_i e_i^c h + \mu \bar{\Xi} \Xi + \epsilon_{i} \bar{\Xi} \ell_i + \tilde y_i \Xi  e_i^c h + H.c. \label{eq:mixinglagrangian} 
\eeq
Prior to electroweak symmetry breaking, we can do a rotation in $(\ell_i  \, \Xi)$ space to remove the $\epsilon_i$ terms at the cost of modifying the $\tilde y_i$ terms. Subsequent rotations on the $\ell_i$ and $e_i^c$ allow diagonalization of the $3\times 3$ submatrix.
After electroweak symmetry breaking and performing the above rotations, the charged and neutral fermion mass matrices are
\beq
\mathbf{M_{charged}} = 
\left( 
\begin{array}{cccc}
m_e &0 & 0 &0 \\
0& m_\mu &0 &0 \\
0& 0& m_\tau &0 \\
\tilde m_e & \tilde m_\mu & \tilde m_\tau      & \mu 
\end{array}
\right), \hskip 1 in \mathbf{M_{neutral}} = 
\left( 
\begin{array}{c}
0\\
0\\
0\\
\mu 
\end{array}
\right).
\eeq
Physical neutrino masses are sufficiently small that we can neglect them.
There is no {\em a priori} reason why $\tilde m_e \le \tilde m_\mu, \tilde m_\tau$. In anarchical models\cite{anarchy}, they may all be the same, or should $\Xi$ carry a lepton flavor charge, it could even be that $\tilde m_e \gg \tilde m_\mu, \tilde m_\tau$.  For future discussion, it is useful to introduce the small parameter $\delta_{i} \equiv \tilde m_{i}/\mu$. 

\subsection{Decays of the heavy leptons}
Let us begin by focusing on the phenomenology of the neutral sector.  All neutral states (both Standard Model and the new heavy state) have identical couplings to the gauge bosons.  So, when the mass matrix is diagonalized, no off-diagonal couplings to the $Z$ boson are created --- the mathematics is identical to that which ensures the absence of FCNCs in the Standard Model. 
However, the differences in diagonalizing the charged lepton mass matrices {\em will} lead to off-diagonal charged current decays. Thus, neutral $\Xi$ states decay to a $W^\pm \ell^\mp$ final state.

The decay modes of $\Xi^{\pm}$ are more complicated.  We assume that the $\Xi^{\pm}$ state is heavier than the $\Xi^{0}$ state.  The charged state has non-vanishing couplings to three final states: $\Xi^{0} W^{\pm \ast}$, $Z \ell^{\pm}$, and $W^{\pm} \nu$.  Typically, the most suppressed of these decays 
is the decay to $W^{\pm} \nu$.  This heavy--light charged current coupling is down by  a factor $(m_{\ell}/\mu) \delta_{i}$, where $m_{\ell}$ is the relevant charged lepton mass.    Were $m_{\ell}$ to vanish, then the rotation on the left-handed doublet has no effect (just as in the neutrino case above).  This is the origin of the suppression.  Thus, the two decay modes of potential phenomenological interest are $\Xi^{0} W^{\ast \, \pm}$ and $Z \ell^{\pm}$.  The first of these receives no suppression from small couplings, but receives a large phase space suppression.  If no additional couplings are added to the theory, the dominant contribution to the $\Delta m_{\Xi} \equiv m_{\Xi^{\pm}} - m_{\Xi^{0}} $ mass splitting is from the loop effect induced by the Coulomb interaction, analogous to $\pi^\pm - \pi^{0}$ mass splitting.  Because this effect is cut-off by the scale of electroweak symmetry breaking, it is finite and is approximately given by $\Delta m_{\Xi} \approx \frac{\alpha M_{Z}}{2} \approx$ 350 MeV.   There is an additional (subdominant) contribution to the splitting between the charged and neutral states of size $\delta^{2} \mu$ due to the mixing with the Standard Model states.  We will see later that  $\delta$ is constrained to be $\lesssim 10^{-2}$.  For the tiny splittings induced by the Coulomb interactions, decays will proceed to $\Xi^{0}$ and a soft charged pion with a rate given by \cite{Thomas:1998wy}
\begin{equation}
\label{eqn:thomaswells}
\Gamma_{\pi \Xi^{0}} = G_{F}^{2}  \cos ^{2} \theta_{C} f_{\pi}^{2}  (\Delta m_{\Xi})^3
\sqrt{1 - \left(\frac{m_{\pi}}{\Delta m_{\Xi}}\right)^2}.
\end{equation}
The remaining decay, $\Xi^{\pm} \rightarrow Z \ell^{\pm}$, is unsuppressed kinematically, but the relevant coupling is proportional to $\delta$.  It is given by
\begin{equation}
\label{eqn:XitoZell}
\Gamma_{\ell Z^{0}} = \frac{m_{\Xi} \delta^{2} g^2}{64 \pi \cos^{2} \theta_{W}} \left(1-\left(\frac{m_{Z}}{m_{\Xi}}\right)^{2}\right)^{2}\left(2+\left(\frac{m_{\Xi}}{m_{Z}}\right)^{2}\right).
\end{equation}
For $\delta > 10^{-7}$, the kinematic suppression of the charged current decay is more severe, and the decay via $\Xi^{\pm} \rightarrow Z \ell^{\pm}$ dominates.   In the remainder of the paper, we will assume that $\delta$ is sufficiently large so that $\Xi^{\pm}$ decays in this way.  

\subsection{Spectrum of annihilation products}
We first consider the case when the dark matter dominantly annihilates into two heavy states.  It is natural to assume the dark matter annihilates equally into $\Xi^{+} \Xi^{-}$ and $\Xi^{0} \Xi^{0}$.  Based on the above discussion, these processes yield $\ell^{+} \ell^{-} ZZ$ and $\ell^{+} \ell^{-} WW$ final states. 
In principle, the mixing of the new vectorlike state can be with any Standard Model lepton family.  Not surprisingly, the case where mixing is dominantly into electrons provides the most favorable to fit the PAMELA positron spectrum: it  produces the hardest positron spectrum possible for a certain amount of antiprotons.  Mixing with $\mu$ and $\tau$ leptons 
will require larger boost factors to produce the positron signal and will result in some tension with the anti-proton data.  

We now address the spectrum that results from dark matter annihilation.  It depends on the masses $m_{\chi}$ and $m_{\Xi}$. 
In the limit $m_{\Xi} \to m_\chi$, the spectrum decomposes into two components: gauge bosons of energy $\sim m_\chi/2$ (assuming $m_\Xi \approx m_\chi \gg m_V$) and electrons and positrons of energy $m_\chi/2$.  Here $m_{V}$ is the mass of the relevant gauge boson.  In the opposite limit, $m_\chi \to m_V $, the energy of the positron from the $\Xi$ decay goes to zero.  The most relevant positrons come directly from gauge boson decay will have maximum energy $m_\chi$.  Finally, one finds the hardest electron/positron spectrum for the lepton from $\Xi$ decay comes for 
\beq
m_{\Xi} = \sqrt{2 m_{\chi} m_V - m_V^2}.
\eeq
The spectra for these three limits are shown in Fig.~\ref{fig:spectra}.

\begin{figure}
  \begin{center}
    \scalebox{0.6}{\includegraphics{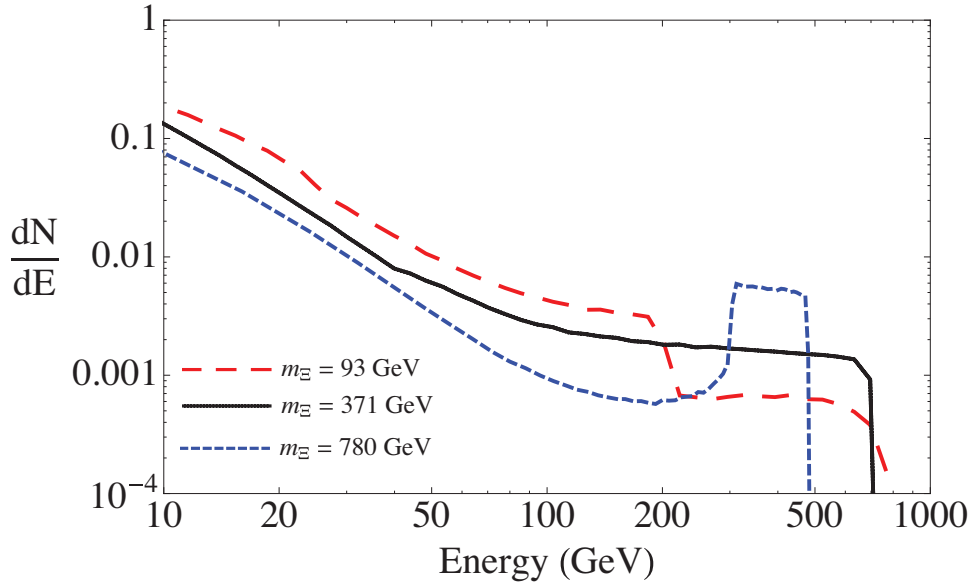}}
  \end{center}
\vspace{-.5cm}
  \caption{Positron spectra from the process $\chi \chi \rightarrow \Xi \bar \Xi$ for three different $\Xi$ masses for a dark matter mass $m_\chi$ = 800 GeV.  These injection spectra are shown prior to propagation and correspond to a single annihilation.  It is assumed that the $\Xi$ decay via a mixing with electrons.  \label{fig:spectra}}
\end{figure}

The addition of two hard leptons to every annihilation allows a fit to the observed positron fraction with a smaller boost factor than would be required if annihilation proceeded directly to gauge bosons.  This provides a relative suppression of  the antiproton contribution. This is roughly an order of magnitude change with respect to the case of pure annihilation to gauge bosons.  Residual $\bar{p}$'s will still provide a constraint. We will explore this in section \ref{sec:pamela}.

Next, we consider the case where dark matter annihilates into  one heavy and one light lepton.  Similar final states occur in Randall-Sundrum theories of dark matter \cite{Ponton:2008zv}.  Again, the natural assumption is that dark matter annihilation populates the charged and neutral states equally.  As long as $m_{\Xi} \lesssim 2 m_\chi$, the $\Xi$ final state is available.  In this limit ($m_{\Xi} \rightarrow 2 m_{\chi}$), no kinetic energy will be available for the $\Xi \ell$ final state.  Thus, the monochromatic light lepton is soft.   Effectively, the final state is a $\Xi$ field at rest, which then decays as described in the previous section.  In the opposite limit of a very light $\Xi$, there is a combination of a hard monochromatic lepton plus a $\Xi$ decay spectrum. We plot the positron injection spectrum in three cases, $m_{\Xi} = 95$ GeV, $\approx m_{\chi}$, and  $\approx 2 m_{\chi}$ in Fig.~\ref{fig:injectionHL} for the case where the final state lepton is an electron.  In this heavy--light case, a monochromatic neutrino is also produced.  We will discuss its consequences in Section IV.   
\begin{figure}
\begin{center}
  \scalebox{0.6}{\includegraphics{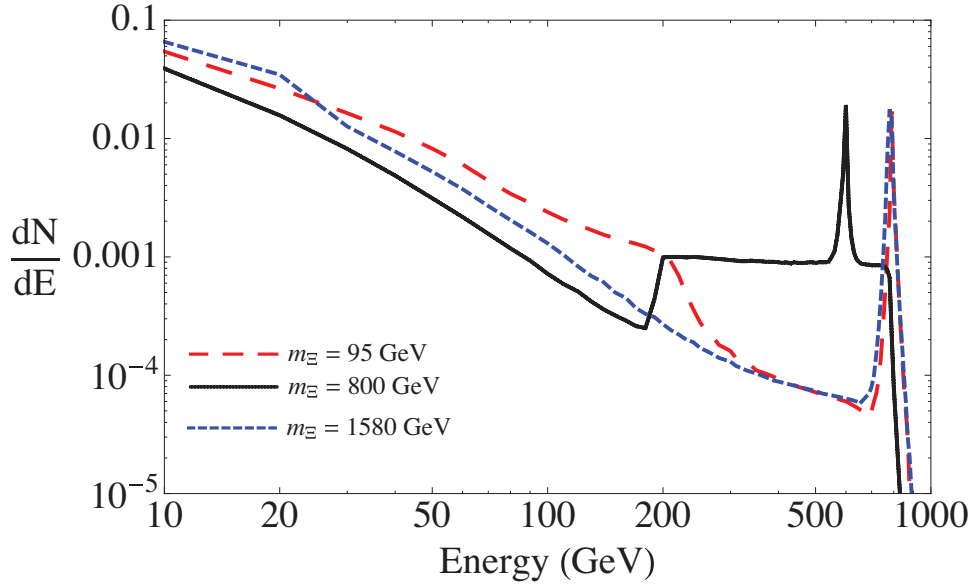}} \hskip 0.5in
  \end{center}
\vspace{-.5cm}
  \caption{Positron spectra from the processes $\chi \chi \rightarrow \Xi e$  
for a dark matter mass $m_\chi$ = 800 GeV and three different $\Xi$ masses.  These injection spectra are shown prior to propagation, and correspond to a single annhilation.    \label{fig:injectionHL}}
\end{figure}

\subsection{Precision Constraints \label{sec:precisionconstraints}}
Although we introduce new states that mix with Standard Model leptons, it is obvious from the outset that there need be no conflict with precision observables. If the mixing with the Standard Model (as controlled by $\delta= \tilde{m}/\mu$) is sufficiently small, there will be no measurable effect on, {\em e.g., $g-2$}, although $\Xi$ will still decay on (cosmologically) short timescales. However, to understand the phenomenology of the heavy states, it is important to consider how large $\delta$ can be.

In the case of a vectorlike doublet, there are two relevant classes of constraints.  First, if both $\delta_{\mu}$ and $\delta_{e}$ are non-zero, a bound arises from the process $\mu \rightarrow 3e$.  Consistency with the non-observation of this process $BR(\mu \rightarrow 3e ) < 10^{-12}$ \cite{Bellgardt:1987du} forces $ \sqrt{ \delta_{e} \delta_{\mu} } < 2 \times 10^{-3}$.  A weaker constraint arises on this combination from $\mu \rightarrow e \gamma$. Second, there are constraints present even if only one $\delta$ is non-zero.  In particular, there are precision electroweak constraints from considering, e.g., the universality of $\Gamma(Z  \rightarrow \ell \ell)$.  This forces $\delta_{\ell} < 2 \times 10^{-2}$

Should we choose the vectorlike state to instead have the quantum numbers of a right-handed electron, there is also an important bound that comes from the universality of $G_{F}$.  In that model, the $SU(2)_{L}$--singlet exotic is mixed with a standard model lepton and effectively reduces the coupling of this light state to the $W$.  Thus, for the singlet model there is an additional constraint, $\delta_{\ell} < 4 \times 10^{-3}$.

So, there is a large window (roughly   $10^{-7}< \delta < 10^{-2}$), where the decay phenomenology described in the previous section applies and is consistent with all precision constraints.  We assume that we are in this window and now move to a detailed discussion of the astrophysical signals of this model.

\section{Explaining the PAMELA results \label{sec:pamela}}

While we have presented a simple scheme for generating hard positrons with few antiprotons, it remains to be seen whether we can fit all of the existing data, and what the minimal boost factor above the thermal cross section is required to reproduce a good fit. 

To investigate the annihilation spectra, we implemented the processes described in section \ref{sec:annihilations} into MadGraph \cite{Maltoni:2002qb}.  This allowed the simulation of the annihilation and decays shown in Fig.~\ref{fig:DMannihilation}. Subsequent decays of the $W$ and $Z$ bosons and hadronization where done with PYTHIA \cite{Sjostrand:2006za}.  This produced an injection spectrum of positrons, electrons, antiprotons, and gamma rays.  This spectrum was input to GALPROP \cite{Strong:1998pw} to propagate the decay products through the galaxy.  For concreteness, an NFW profile was used \cite{Navarro:1996gj}
\begin{equation}
\rho(r) = \rho_{\odot} \left(\frac{r_{\odot}}{r} \right) \left(\frac{1+\left(r_{\odot}/r_s \right)}{1+\left(r/r_s\right)} \right)^2,
\end{equation} 
with $r_s = 20$ kpc, where $r_{\odot} = 8.5$ kpc is the galactocentric distance of the sun and $\rho_{\odot} = 0.3$ GeV/cm$^3$ is the local dark matter density, although the shape of the profile will have little effect on the observed positron fraction.  For positron, electron, proton, and antiproton flux backgrounds, we use the model described in \cite{Cholis:2008vb} with the Alfv\'en velocity $v_A = 20$ km/s.  The diffusion coefficient is taken to be $D = \beta (5.88 \times 10^{28}$  cm$^2$/s) (R/4 GV)$^{1/3}$, where $\beta = v/c$ and $R$ is the rigidity.  The height of halo region is set to $L = 4$ kpc. 

There are a number uncertainties that enter into the calculation of the rates. In addition to the cross section itself, indirect detection signals are proportional to the number density squared of dark matter particles. The uncertainties associated with this are usually encoded into a ``boost factor''.  We  define the boost factor as 
\beq
BF = \frac{1}{V_{CR}} \left( \int d^3 x \frac{n_{true}^2(r)}{n^2(r)}\frac{\langle \sigma v \rangle}{\langle \sigma v \rangle_{thermal}} \right).
\eeq
Here $ n(r) = \rho(r)/m_\chi$ is the previously mentioned NFW profile with local density $\rho_0=0.3\, \rm{GeV \; cm^{-3}}$. $n_{true}(r) = \rho_{true}(r)/m_\chi$ is the actual (possibly clumpy) number density of the dark matter particles in the halo, and the integration is over a region $V_{CR}$ defined by cosmic ray propagation.  

The integral over the number density contains all of the information on astrophysics and is subject to both uncertainty in the halo profile and uncertainties on propagation.  Clumpiness gives an increase in $n^2$ when compared to a  smooth profile.   Should the boost factor arise from astrophysics, it is expected that the boost factors could be different for positrons and antiprotons.  Depending on the positions of subhalos with respect to the solar position, it could lead $BF_{e^+} / BF_{\bar{p}} \sim 3$ \cite{Lavalle:1900wn} and energy dependent boost factors.  We will ignore the possibility of energy dependent boost factors in this work, but will bear in mind the possibility that astrophysical uncertainties might somewhat ameliorate tensions between positrons and anti-protons.  Note that uncertainties in the local number density could be a factor of $\sim 2$  (yielding a boost of $\sim 4$) \cite{Bergstrom:2000pn, PDBook}. Altogether, boosts of up to $\sim 10$ arising from astrophysics alone are likely reasonable, while much higher boosts (as we shall find are necessary for heavier WIMPs) likely would rely on a contribution from a non-trivial dark matter cosmology.

It should also be noted that uncertainties in the propagation of antiprotons could suppress the dark matter contribution to the antiproton signal by up to an order of magnitude \cite{Donato:2008jk}.  The effects of these changes on the positron flux is more modest.  
So, while in this work we quote a boost factor for the positron signal and naively apply the same the boost factors to the anti-protons, it should be kept in mind
that masses that initially appear to be ruled out by the observed $\bar{p}/p$ ratio might be allowed once a full accounting of  these uncertainties is made.

We show the results of our analysis in Fig.~\ref{fig:massvary1}. We find that it is possible to fit the PAMELA positron data for a variety of masses.  For each set of masses, we find the boost factor (as defined above) that gives the best fit to the data.
To avoid the complications  of solar modulation \cite{Potgieter:2001, Clem:1996}, we perform a $\chi^2$ fit to only the four highest energy bins of the PAMELA data (where solar modulation is known to have little effect).  The results are the boost factors shown in Fig.~\ref{fig:massvary1}.  Since PAMELA data were taken during a negative polarity part of the solar cycle, correcting for modulation would reduce positron ratio at lower energies,  bringing the curves into qualitatively better agreement with the data. 

The smallest dark matter mass found to be consistent with the PAMELA positron data is $m_\chi$ = 200 GeV with $m_\Xi \gg m_W$.  In this case, the boost factor required is only 6, which could very plausibly be entirely due to astrophysics.  Thus, dark matter signals could result from a thermal annihilation cross section with no need for non-thermal production or late time cross section enhancements.  As $m_\chi$ increases, the boost factor required is increasingly unlikely to be given by astrophysics alone, and an increase in the annihilation cross section beyond the thermal one is likely necessary.  This could be consistent, for example,  with a cosmological history where the dark matter is produced non-thermally, or if the late-time properties of the dark matter change \cite{PierceCohenMorrissey}. If $\Xi$ decays to muons or tau leptons, this will produce a softer positron spectrum and will require boost factors to fit the positron data.  These will come into more tension with the antiproton data.  For muons, the difference is roughly a factor of two from the electron case.

Dark matter masses $m_{\chi} \sim 500$ GeV are in tension with the antiproton data. Keeping in mind the earlier caveats mentioned on astrophysical uncertainties (both propagation and clumpiness), a prediction of this model is that the dark matter have mass less than 500 GeV.  A turn-over or plateau in the positron fraction should be imminent.  A reasonable fit to both the $\bar p$ and positron data is possible in the case where the $\Xi$ dominantly mixes to muons for $m_{\chi} < 400$ GeV.


\begin{figure}[p]
  \begin{center}
    \scalebox{0.32}{\includegraphics{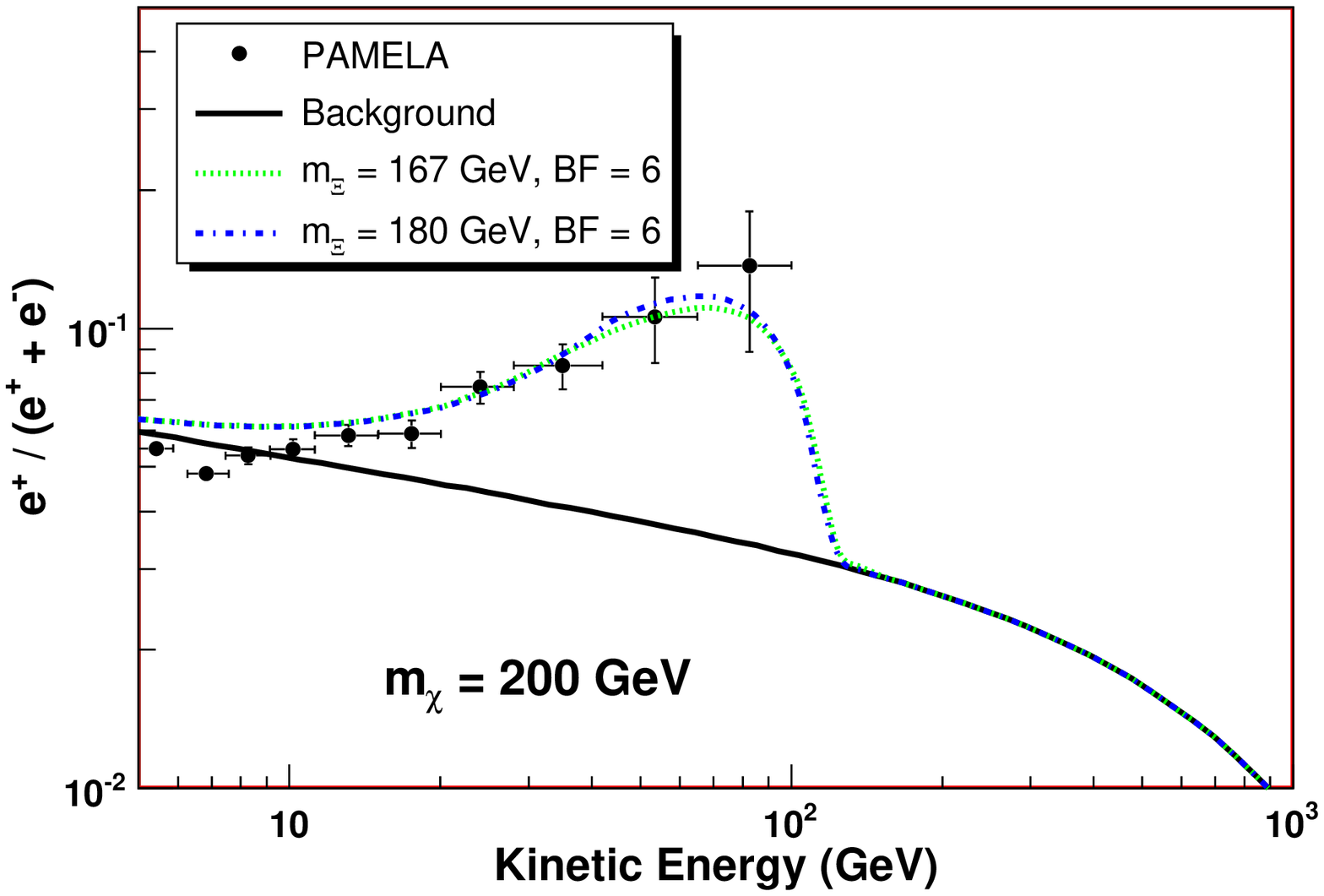}}
    \scalebox{0.32}{\includegraphics{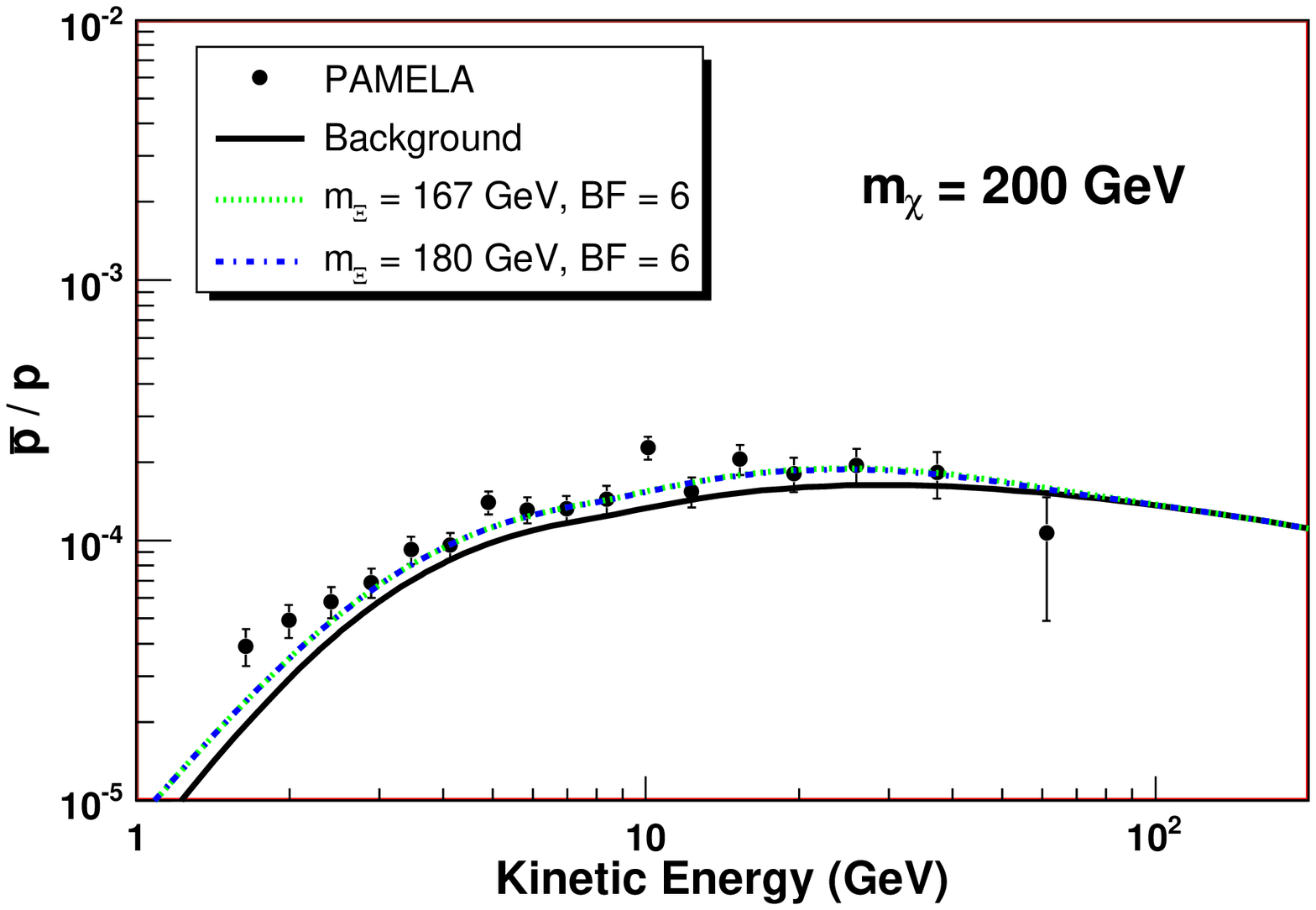}}
    \scalebox{0.32}{\includegraphics{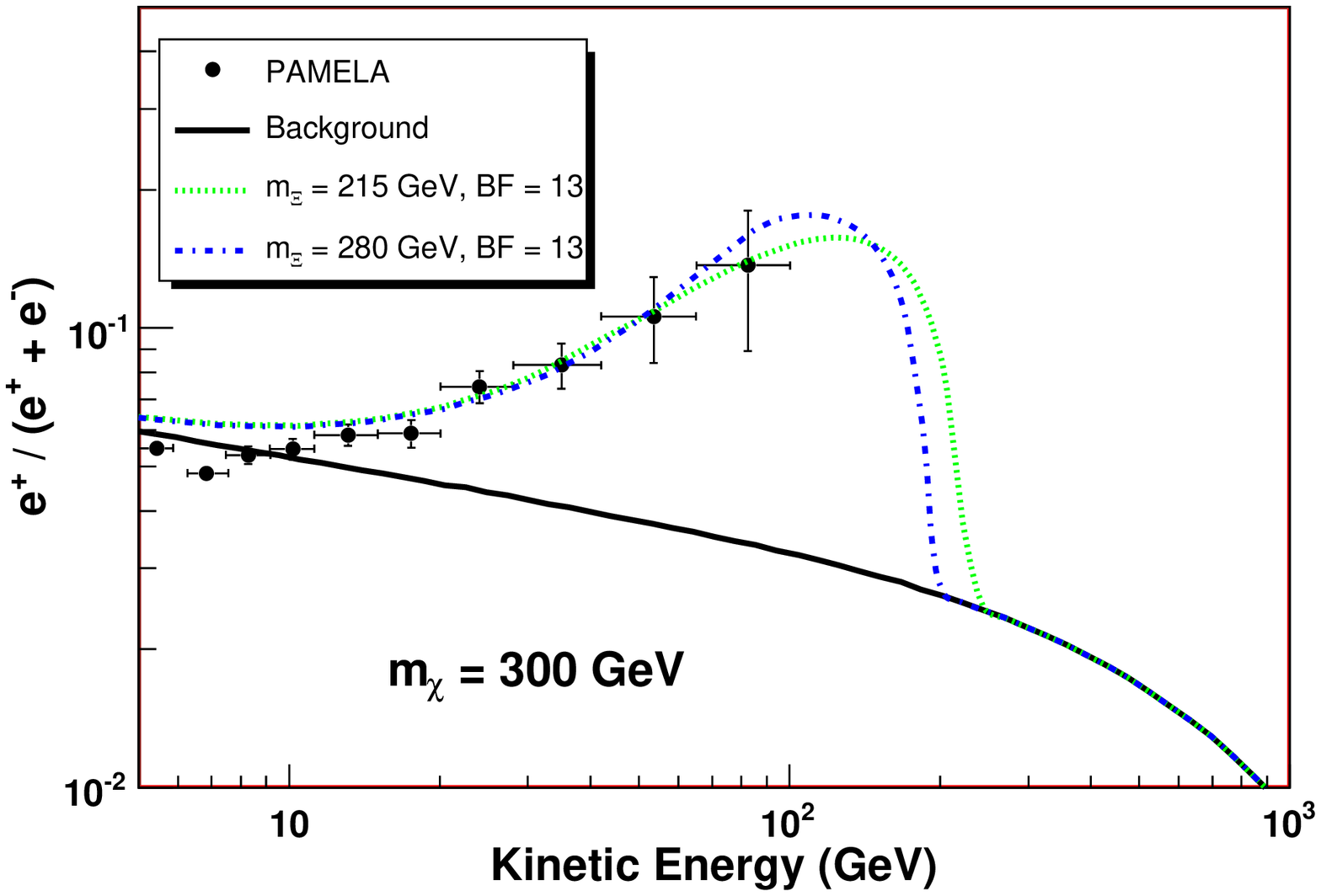}}
    \scalebox{0.32}{\includegraphics{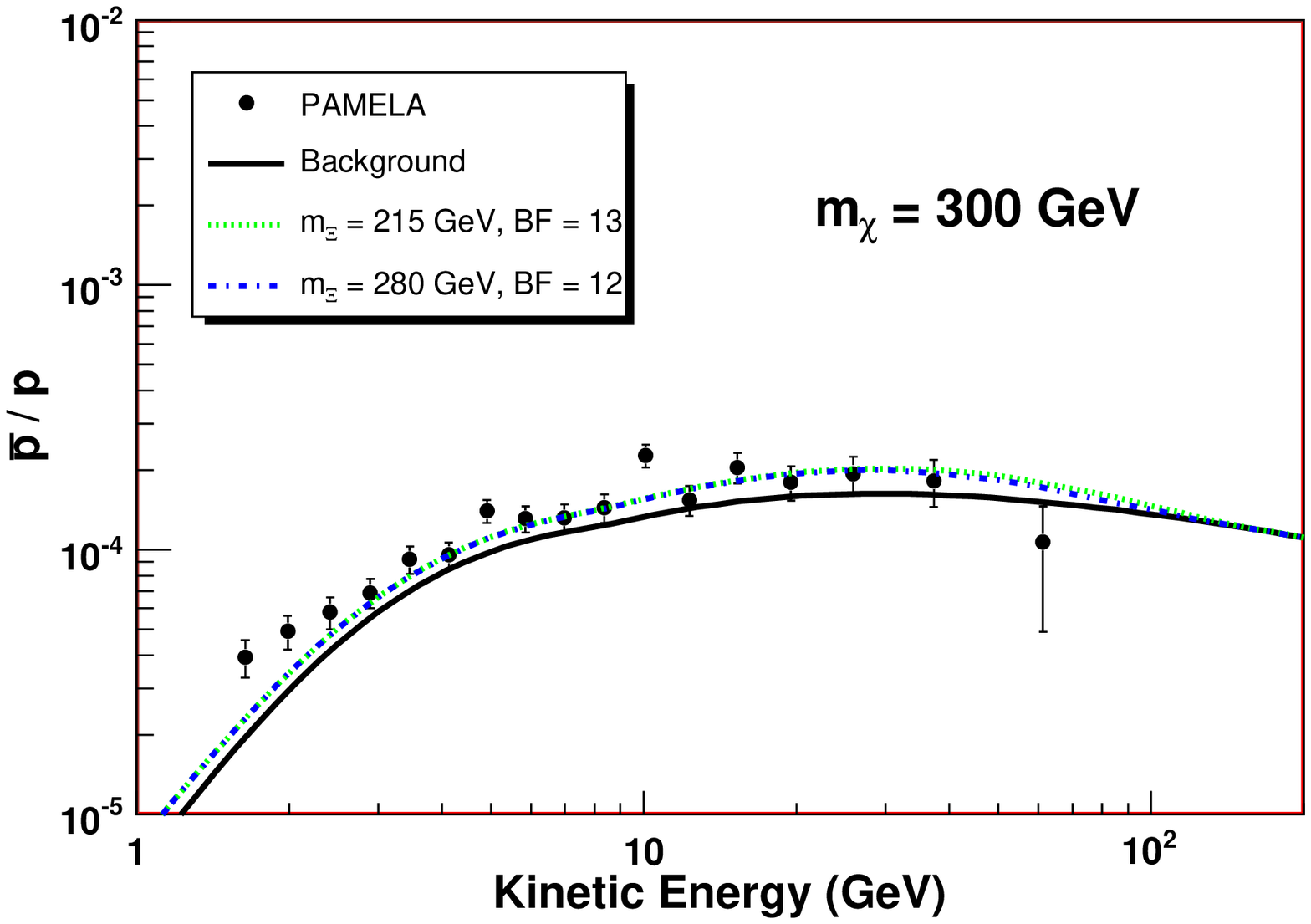}}
    \scalebox{0.32}{\includegraphics{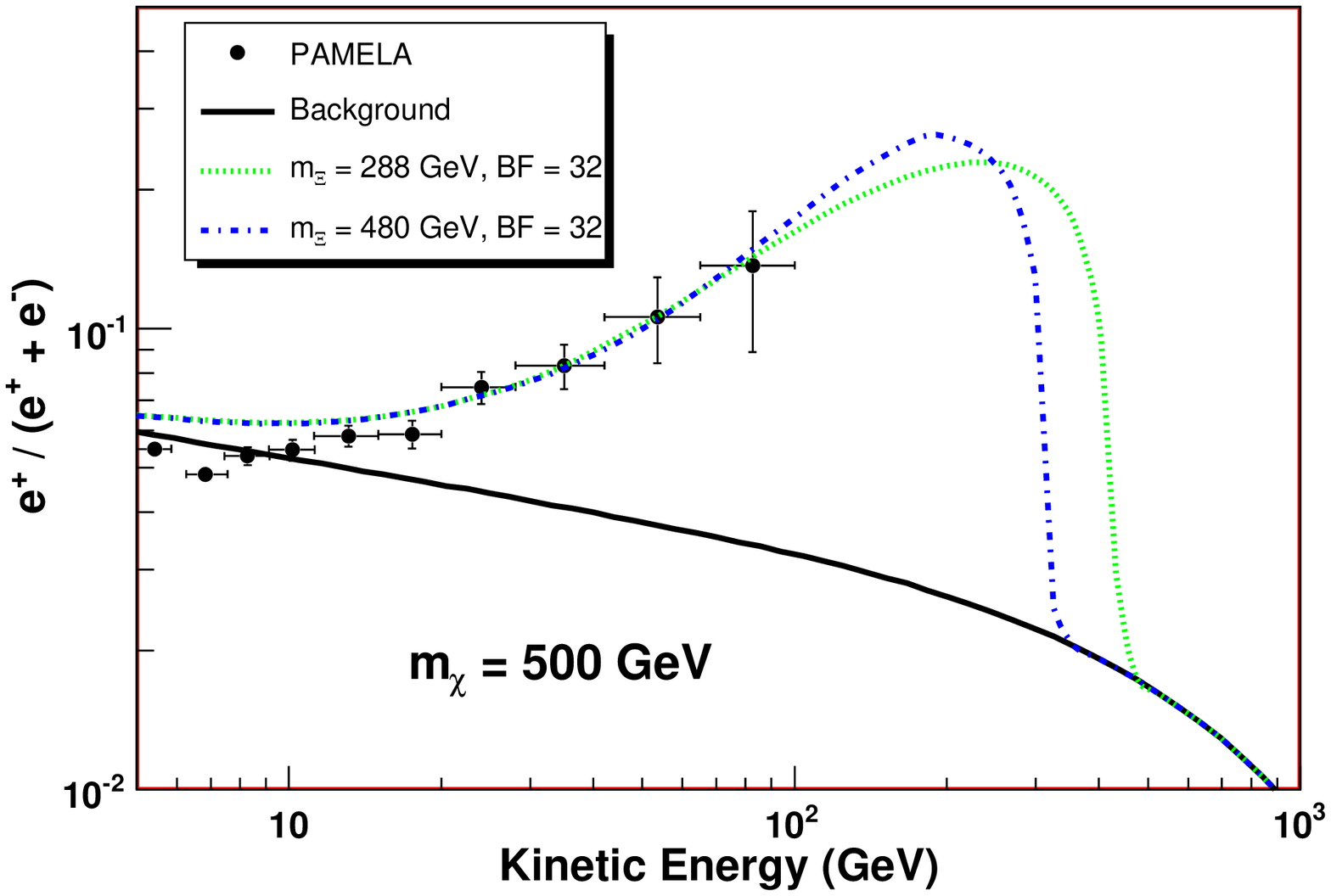}}
    \scalebox{0.32}{\includegraphics{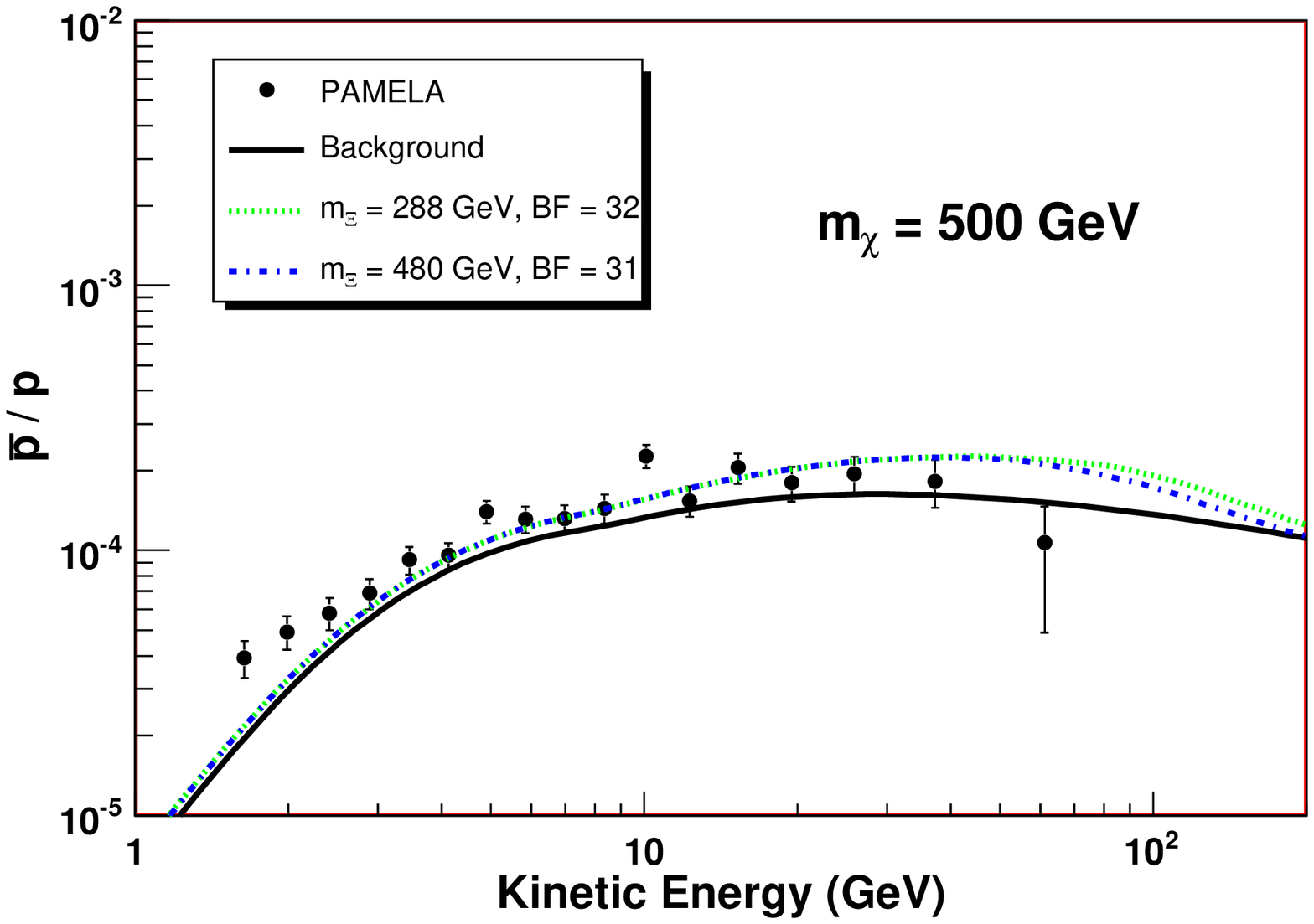}}
    \scalebox{0.32}{\includegraphics{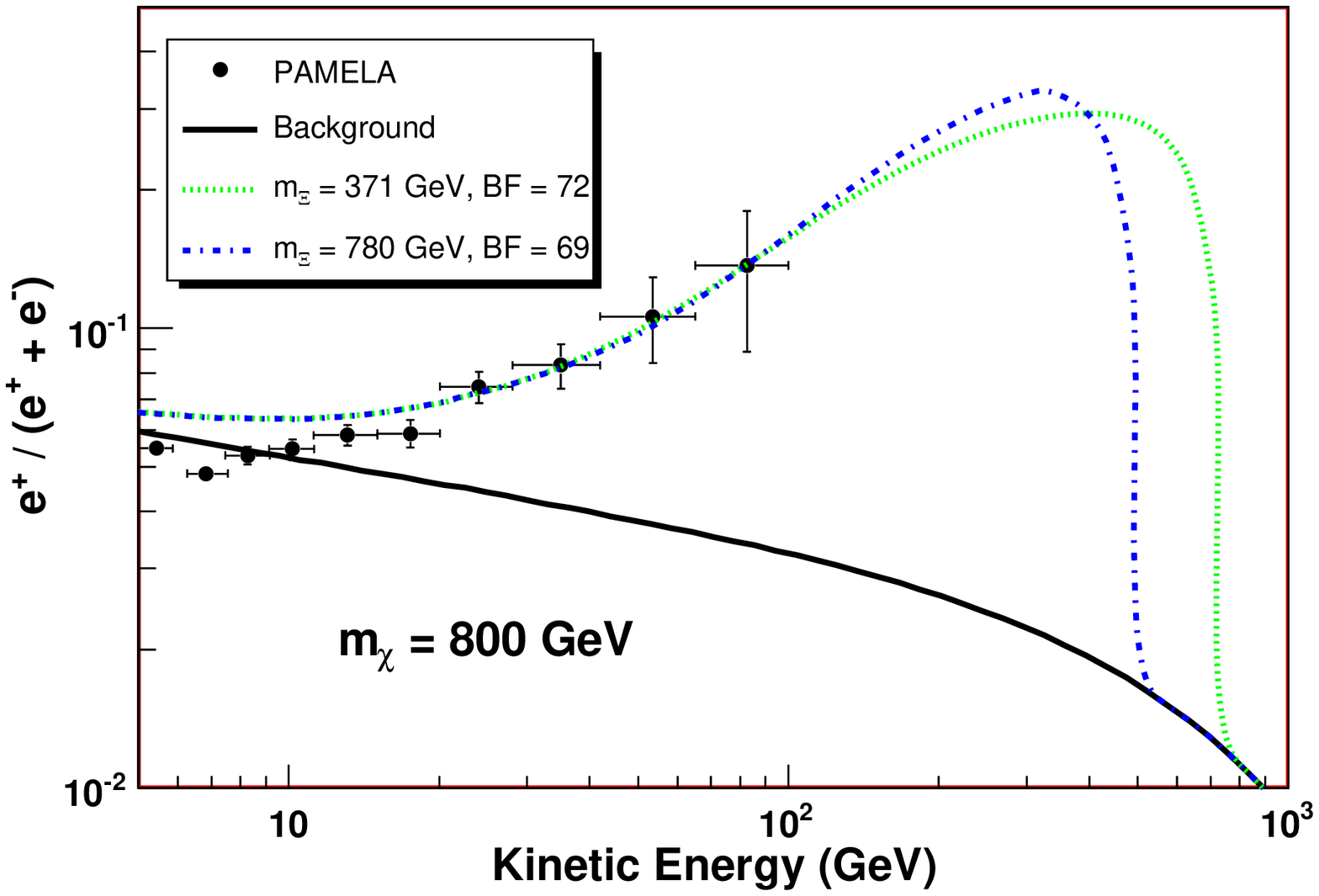}}
    \scalebox{0.32}{\includegraphics{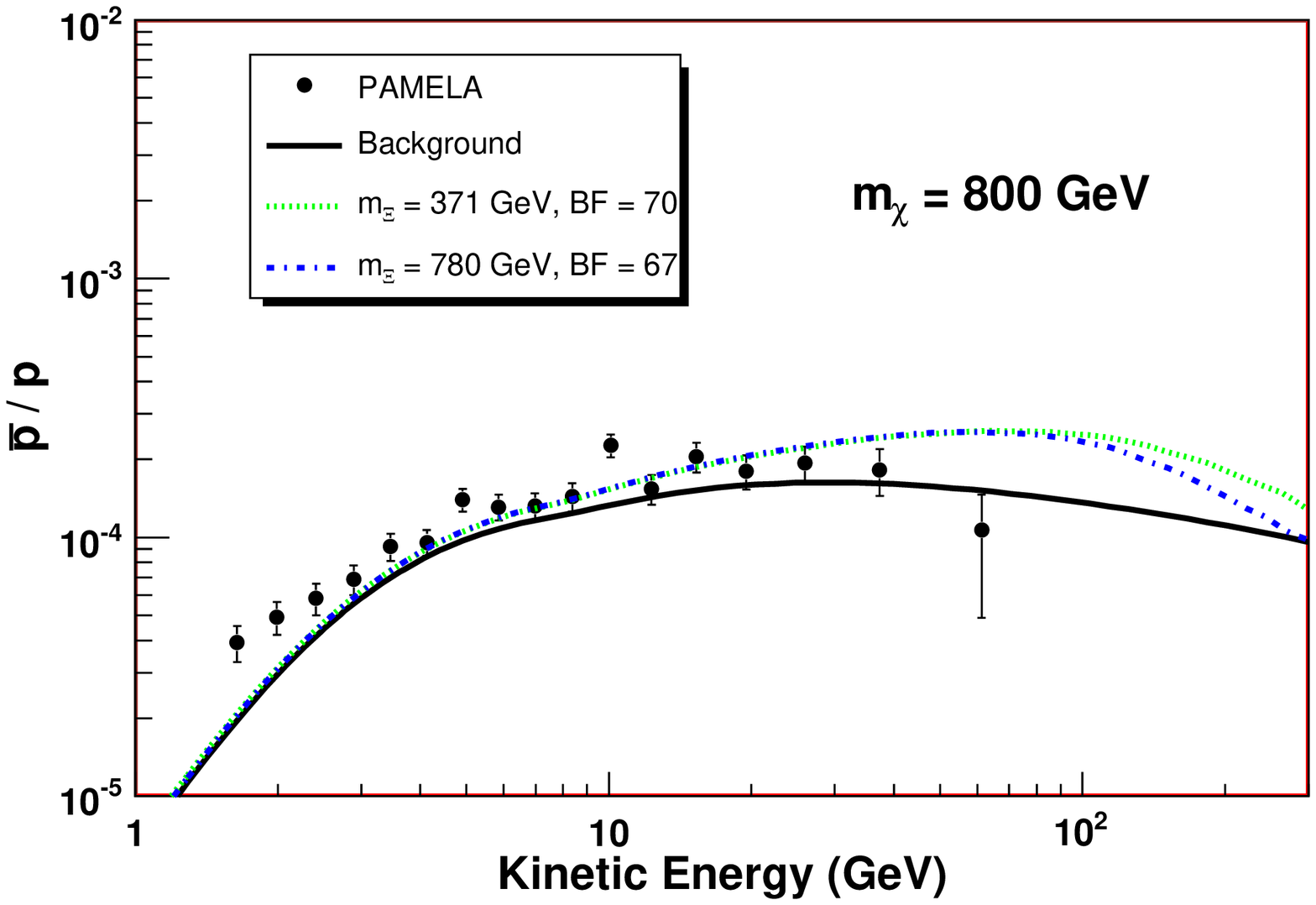}}
  \end{center}
  \vspace{-.4in}
  \caption{Positron fraction and antiproton fractions  $\chi \chi \rightarrow \Xi \bar{\Xi}$.  The boost factors are relative to $3 \times 10^{26}\, $cm$^3$/s, and are found by scaling the positron flux to fit the observed positron fraction. In the $m_{\chi}$ =200, 300 GeV cases, the boost factor required is sufficiently small that the dark matter might have a thermal relic abundance.  \label{fig:massvary1}}
\end{figure}

\begin{figure}[p]
  \begin{center}
    \scalebox{0.32}{\includegraphics{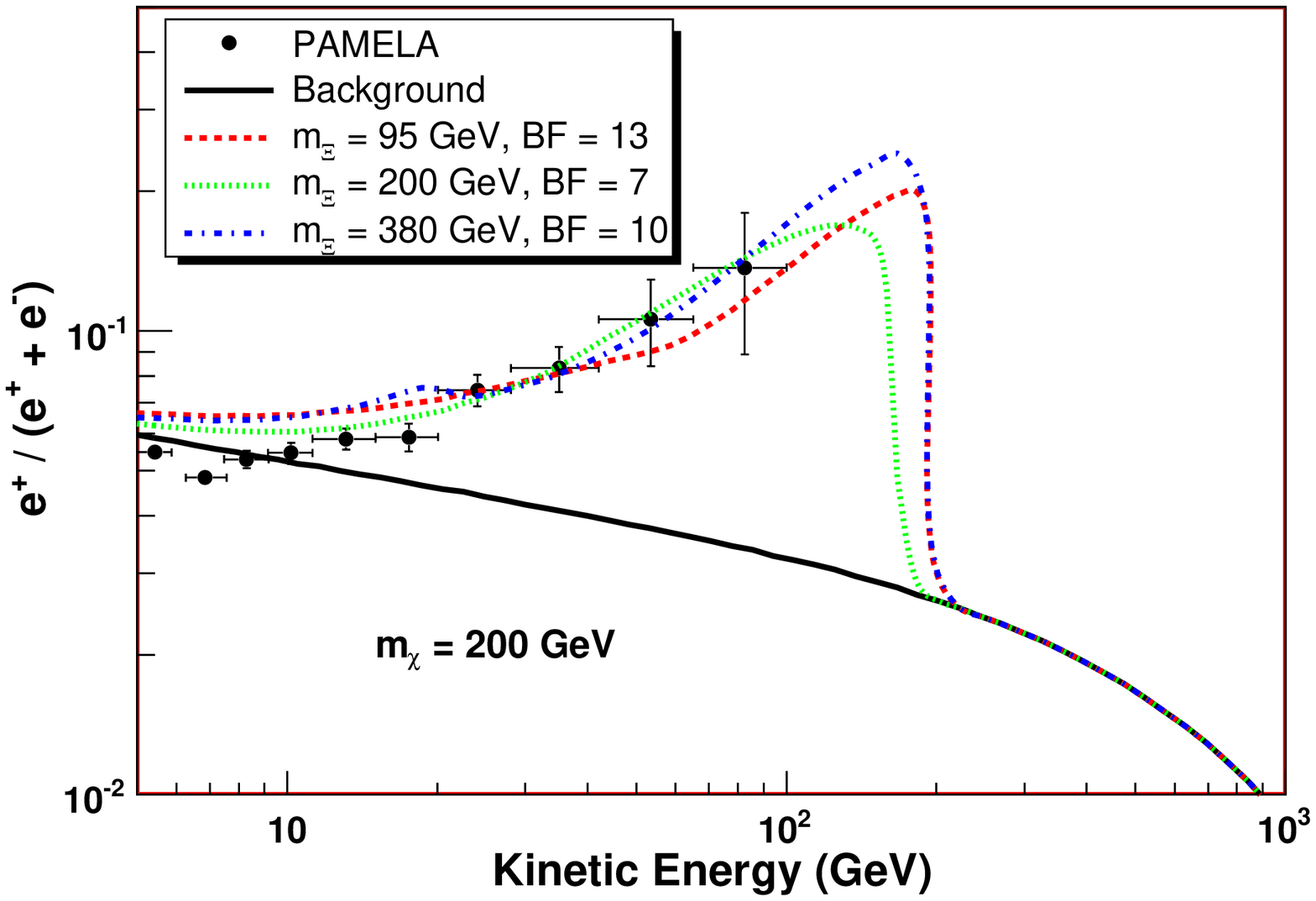}}
    \scalebox{0.32}{\includegraphics{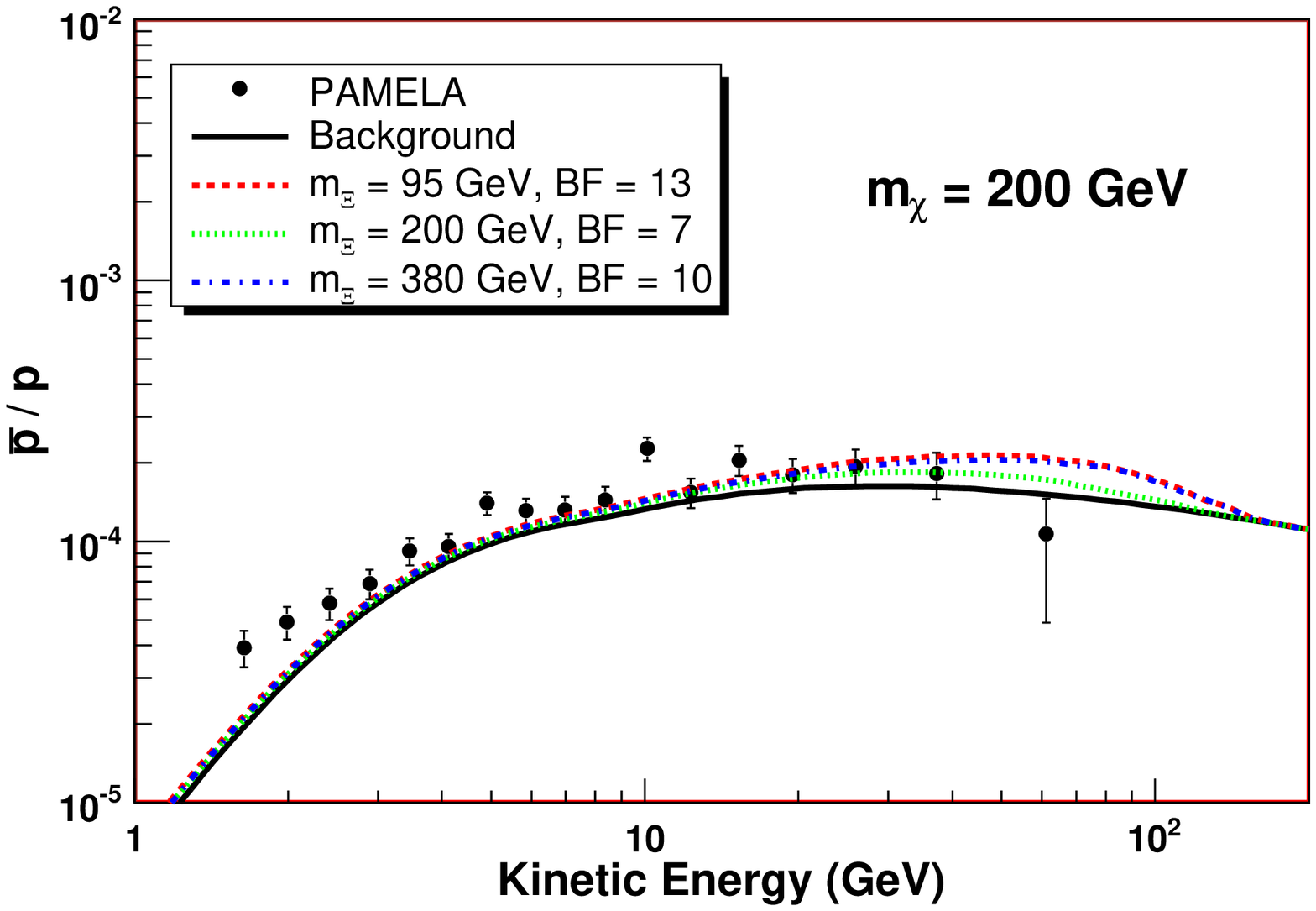}}
    \scalebox{0.32}{\includegraphics{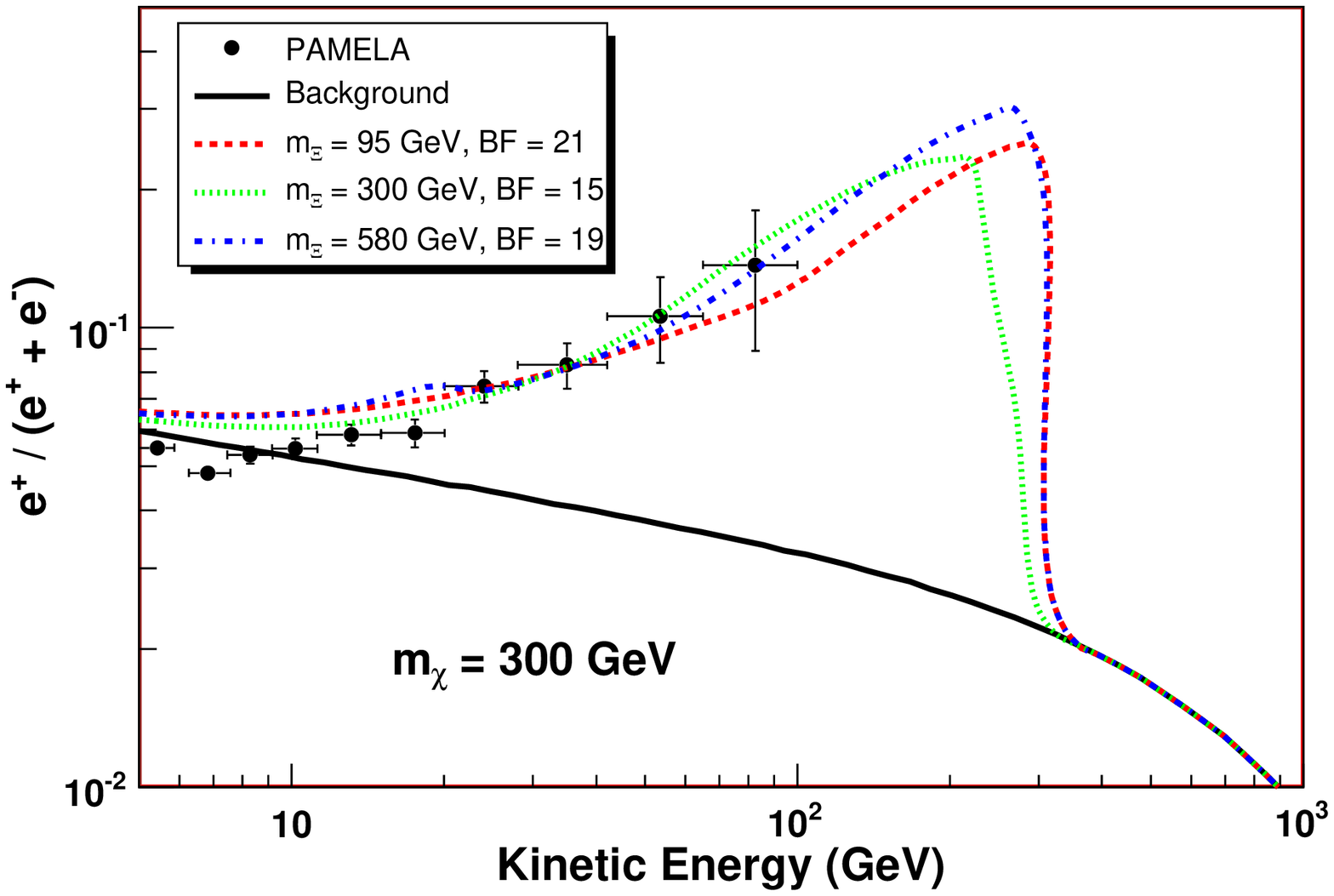}}
    \scalebox{0.32}{\includegraphics{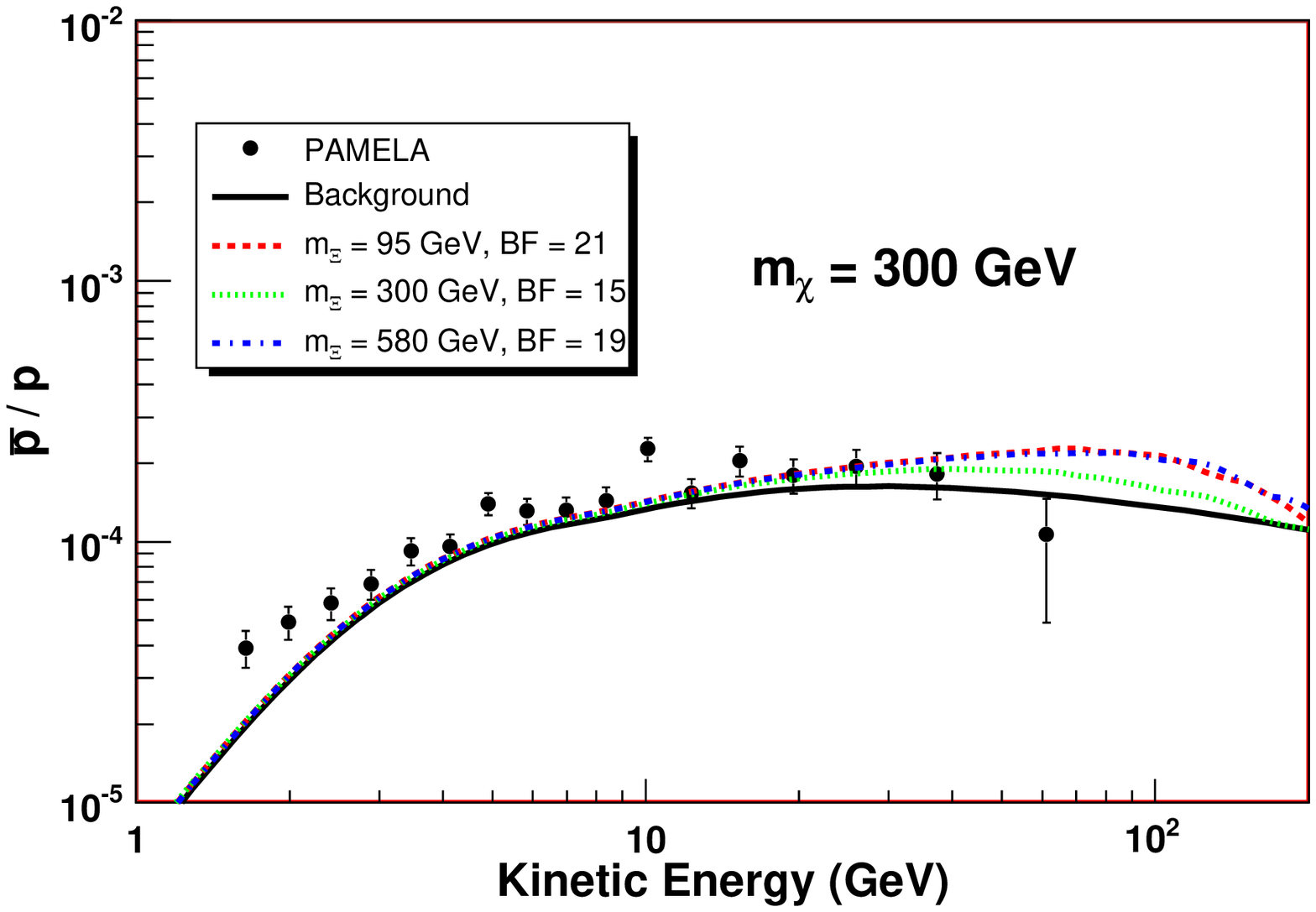}}
    \scalebox{0.32}{\includegraphics{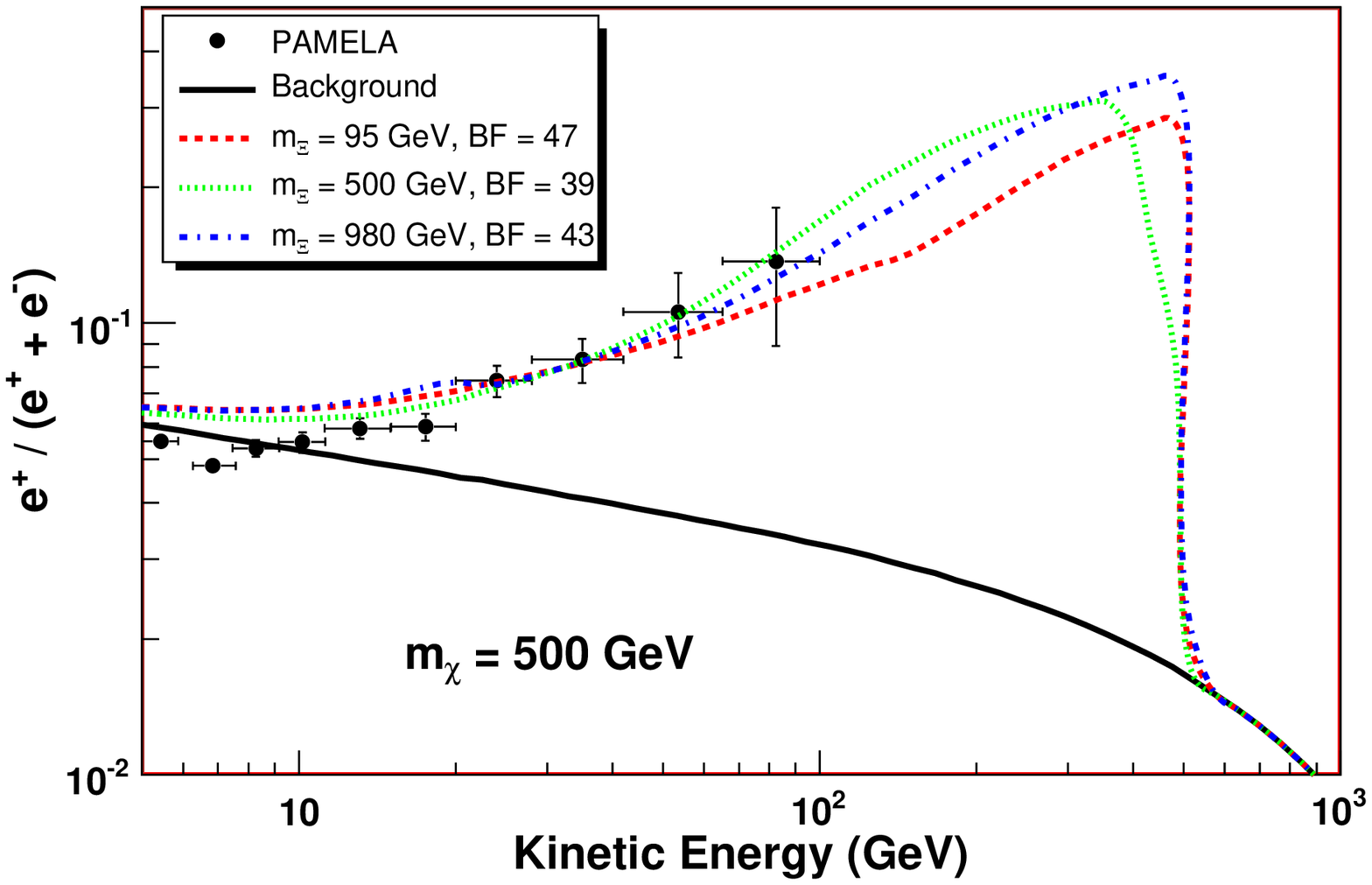}}
    \scalebox{0.32}{\includegraphics{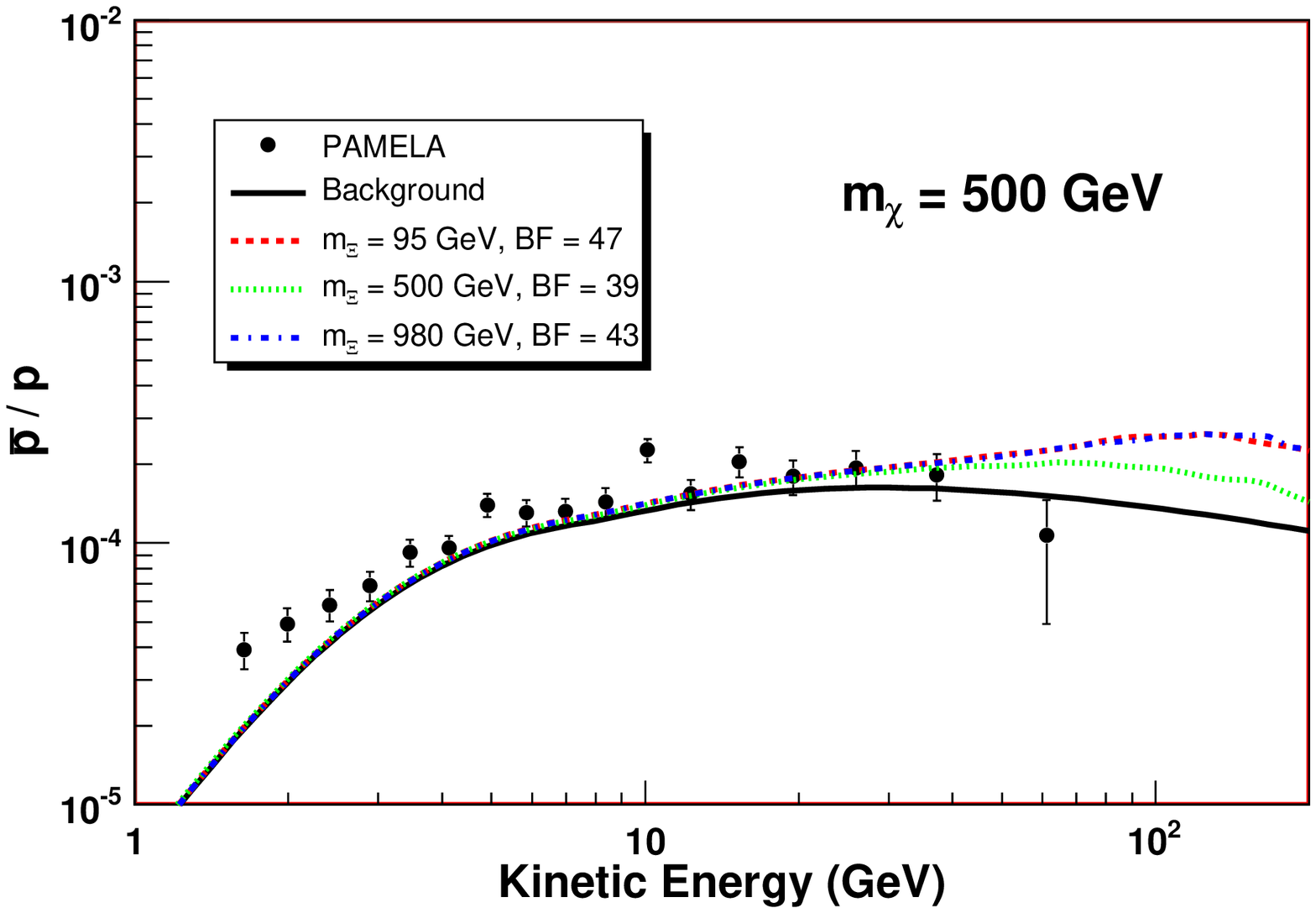}}
    \scalebox{0.32}{\includegraphics{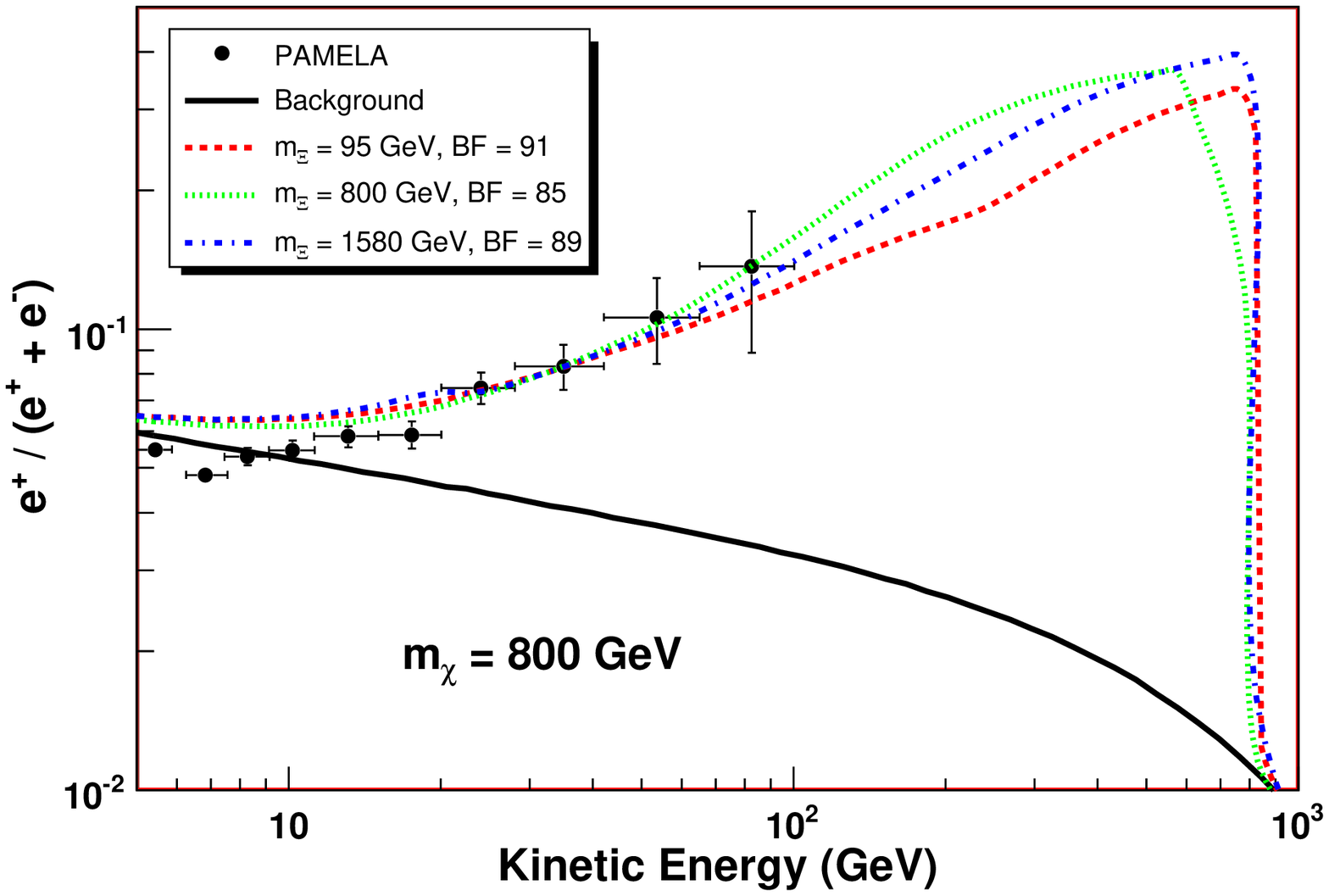}}
    \scalebox{0.32}{\includegraphics{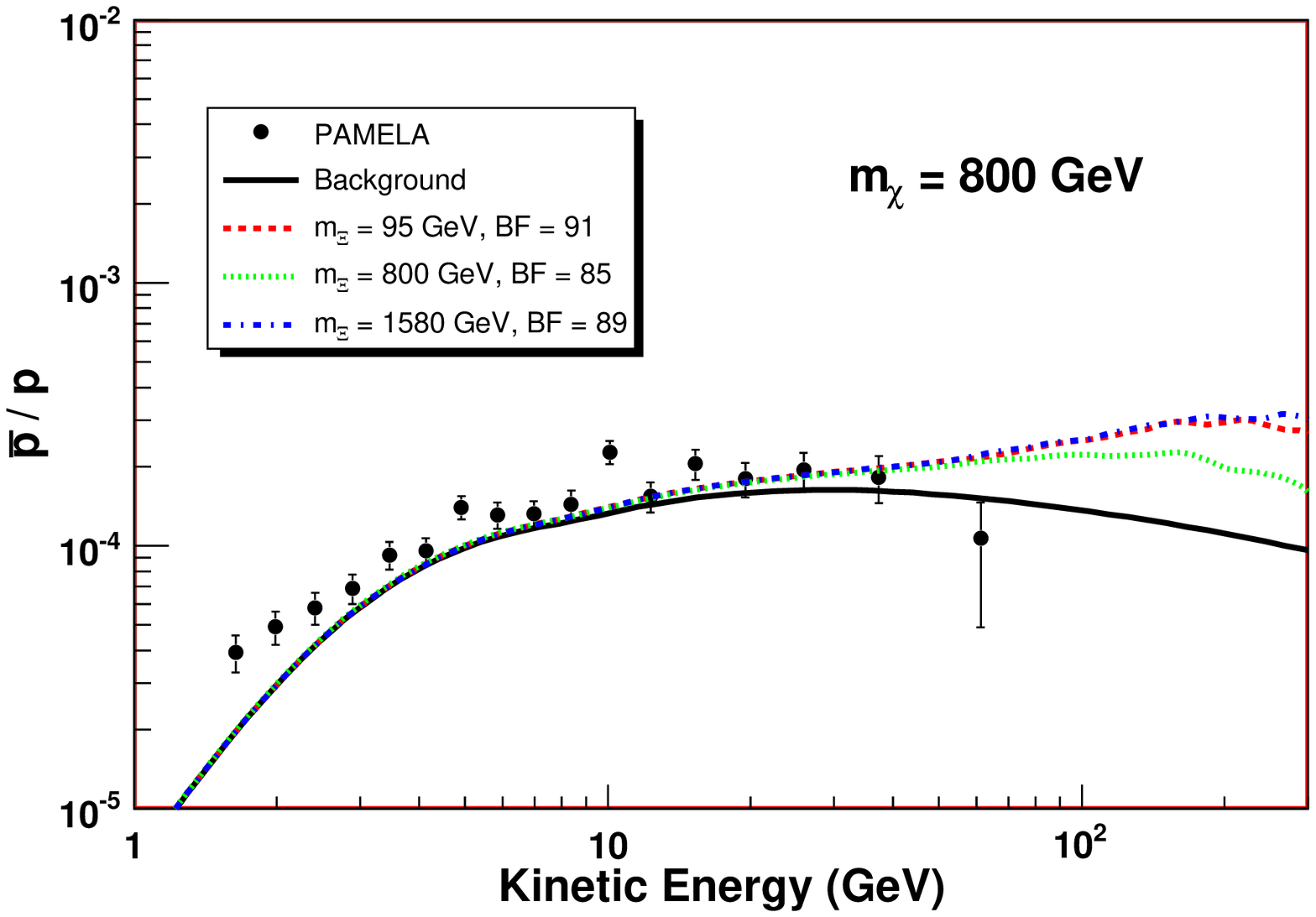}}
  \end{center}
  \vspace{-.4in}
  \caption{Positron fraction and antiproton fractions for $\chi \chi \rightarrow \Xi e$.  The boost factors are relative to $3 \times 10^{26}\, $cm$^3$/s, and are found by scaling the positron flux to ft the observed positron fraction. In the $m_{\chi}$ = 200 GeV case, the boost factor required is sufficiently small that the dark matter might have a thermal relic abundance.  \label{fig:massvary2}}
\end{figure}

When annihilation occurs via $\chi \chi \rightarrow \Xi e$, the positron spectra have slightly cuspier shapes than when annihilation proceeds through $\chi \chi \rightarrow \Xi \bar \Xi$, see Fig.~\ref{fig:massvary2}.  The boosts required are similar to the $\Xi \bar{\Xi}$ case, and the fits are comparably good. Annihilations via $\chi \chi \rightarrow \Xi \mu$ can also give good fits, but require somewhat larger boosts, leading to some tension with anti-proton results, even for lower dark matter masses.  A dark matter mass of $m_{\chi}=200$ GeV that decays in this way is marginally consistent with the positron and antiproton observations for a boost factor of 10-20, depending on the $\Xi$ mass. 

Finally, if $\chi\chi \rightarrow \Xi \bar{\Xi}$ annihilation dominates, a choice of  $m_\chi = 800$ GeV and $m_{\Xi} = 370$ GeV allows a qualitative fit to the shape and position of the ATIC data \cite{:2008zz}, demonstrated in Fig.~\ref{fig:ATIC800}.  In this figure, the background is scaled such that it saturates the low energy data.  While the signal does not reproduce the ATIC data precisely, we find this qualitative fit encouraging.  If the ATIC anomaly persists, and this model is to explain it,  this points to a  $m_\chi = $ 800 GeV to 1 TeV dark matter particle with a large boost factor, $\sim$70.  Non-standard choices for the propagation parameters would be necessary in to reconcile the high energy antiprotons production with the PAMELA data (Fig.~\ref{fig:massvary1}).

\begin{figure}
  \begin{center}
    \scalebox{0.5}{\includegraphics{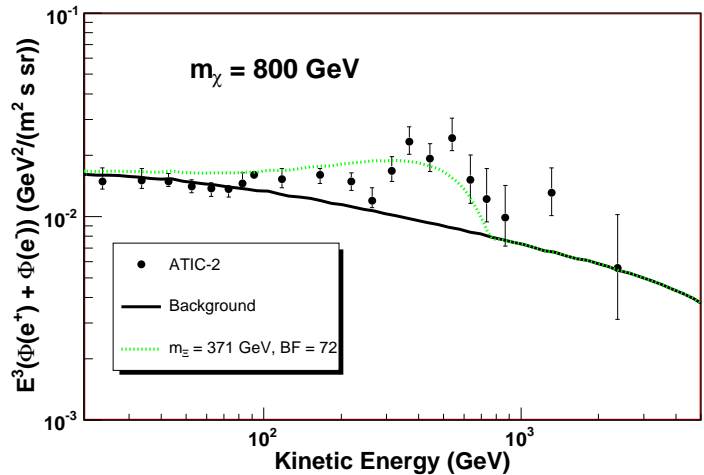}}
  \end{center}
  \caption{ATIC signal for $m_\chi = 800$ GeV and $m_{\Xi} = 371$ GeV. The background has been scaled to saturate the low energy total flux.  The required boost factor is 72, indicating a non-standard cosmological history for the dark matter. \label{fig:ATIC800}}
\end{figure}

\section{Neutrino Signals}

The detectability of neutrino signals, in particular as related to models to explain PAMELA has recently been studied \cite{Beacom:2006tt,Hisano:2008ah,Liu:2008ci}.  Upcoming experiments such as IceCube \cite{Ahrens:2003ix} and ANTARES \cite{Ageron:2008hv} could potentially show sensitivity. 

If annihilation proceeds as $\chi \chi \rightarrow \Xi \bar \Xi$, with $\Xi$ an electroweak doublet, there are essentially no hard neutrinos produced. However, in the case that annihilations proceed $\chi \chi \rightarrow \ell \bar \Xi$, such processes will produce equal numbers of monochromatic charged leptons and neutrinos.  This monochromatic $\nu$ presents an exciting experimental target.

The background from cosmic ray neutrinos \cite{Honda:2006qj} can be well modeled by a power law. Following \cite{Beacom:2006tt}, we examine the signal at a neutrino telescope by requiring a signal to background ratio of 1 in a bin from $10^{-.5} m_\chi$ to $m_\chi$ when compared with the angular average of $\nu_\mu + \bar \nu_\mu$. We assume that oscillations yield $1/3$ of all neutrinos as muon flavor.   The signal is proportional to the line-of-sight integral of the dark matter density averaged over the relevant solid angle, $J_{\Delta \Omega}$. We calculate $J_{\Delta \Omega}$ of 680 for a  $2^{\circ}$ region around the galactic center in an NFW profile.  This solid angle corresponds to the angular resolution of ANTARES.  Taking this gross analysis requiring S:B=1, we find that a detectable signal can be found for boosts of roughly 320\footnote{Here the boost refers specifically to the cross section into $\nu \bar \Xi$ rather than the total cross section.}. 

Such a limit depends only weakly on the dark matter mass (assuming $\mu \ll m_\chi$), although for very large neutrino telescopes where the background statistics are large enough to model the background, the appropriate comparison is naturally $S/\sqrt{B}$ and greater sensitivity would be possible. For reference, in ten years of IceCube running we expect ~200 background events from 500-1000 GeV in a 2$^{\circ}$ region\cite{Liu:2008ci}.  Thus, without a careful examination of background models, we cannot say how detectable this scenario would be at lower masses. For higher masses, detection will be statistics limited, and the S:B criterion is the right one.  This case seems borderline in its detectability.  
Ruling out this model with neutrinos seems difficult. That said, there is substantial evidence within numerical simulations for many subhalos in the Milky Way\cite{Diemand:2006ik,Diemand:2007qr,Springel:2008cc}. Should one of these provide a larger $J_{\Delta \Omega}$, for instance by being cuspy or very nearby, a positive detection remains tantalizingly possible. IceCube has good angular resolution and could focus on one of these regions in an effort to suppress the atmospheric neutrino background.

In fact, if this substructure is sufficiently generous, then it might be possible to observe the $\nu$'s produced in the case where the $\Xi$ is an $SU(2)_{L}$ singlet, but dark matter annihilations go to $\Xi \bar \Xi$.  In this case, the neutrinos have a spectrum very similar to that seen for positrons in Fig.~\ref{fig:spectra}.  They have energy roughly a factor of two lower than the case where direct $\Xi \nu$ dominates.  Due to the steeply falling power law nature of the background, these neutrinos would have to be visible over a background roughly an order of magnitude higher.

Lastly, we should note that the observability of the galactic center signals rely on cuspy profiles into the inner $2^\circ$.   Should those profiles extend into the inner $\sim 0.5^\circ$, strong gamma ray limits from HESS may constrain possible neutrino signals \cite{Mack:2008wu, Bertone:2008xr, Bergstrom:2008ag,Meade:2009rb}.

\section{Collider Signals}
While dark matter signals are often elusive at a collider, we have the exciting possibility of directly producing the new $\Xi$ states that dominate dark matter annihilation.  The cases where the $\Xi$ has quantum numbers of a leptonic doublet or a right-handed electron are interesting.  These states will be produced via the Drell-Yan process.  The collider phenomenology of the case where the $\Xi$ is a complete Standard Model singlet  is uninteresting, simply because it will not be produced at an important level.

First, we consider the doublet case.  Production of $\Xi^+ \Xi^-$, $\Xi^0 \Xi^0$, and $\Xi^{\pm} \Xi^0$ are possible.  The production of $\Xi^+ \Xi^-$ leads to a $ZZ\ell^{+}\ell^{-}$ final state.  The cross section at the LHC for this mode is 27 fb at $m_{\Xi} = 200$ GeV.  In 300 fb$^{-1}$, we expect $\sim 50$ six lepton events, with 4 leptons reconstructing to a pair of $Z$ bosons.  While very few events are expected, they would be spectacular; irreducible physics backgrounds in this channel are vanishingly small. In addition, one could expect observation in the ($ZZ \ell \ell \rightarrow 4 \ell + 2j$) mode  and perhaps even the $2\ell+4j$ mode.  We estimate the background in the $4\ell + 2j$ case to be 9 fb from ALPGEN \cite{Mangano:2002ea} using the process $ZZjj \to 4 \ell jj$.  These backgrounds could be substantially reduced by requiring that the jets reconstruct a $Z$ and applying a $Z$ veto on one of the pairs of leptons. Lepton flavor tagging could also additionally suppress backgrounds. Before cuts, we expect $\sim 500 $ 4$\ell$+2j signal events in 300 fb$^{-1}$ for $m_{\Xi} = 200$ GeV.  Assuming these cuts reduce the background by a factor of a few without compromising the signal, this will give a S:B$\sim$1.  The signal would fall to $\sim 180$ events in 300 fb$^{-1}$ at $m_{\Xi}$ = 300 GeV.   At the Tevatron, the cross section for $\Xi^{+} \Xi^{-}$ production for $m_{\Xi} = 200$ GeV is 1 fb, too small to be seen.  

If the mixing between the $\Xi^{\pm}$ and the Standard Model is very weak, then the $\Xi^{\pm}$ could be long-lived on detector time scales.  However, in the doublet case, the existence of the $\Xi^{\pm} \rightarrow \Xi^{0} \pi^{\pm}$ decay mode unsuppressed by $\delta$ (Eqn.~\ref{eqn:thomaswells}), means that the tracks do not extend much more than a centimeter \cite{Thomas:1998wy}.  This makes observation of these events unlikely.  This changes in the case where the $\Xi^{\pm}$ is a SU(2) singlet.  Then for sufficiently small mixings, a striking long-lived charged track is possible, as all $\Xi$ decays are suppressed by $\delta$.   Depending on the size of $\delta$, there is also a possibility \cite{Jedamzik:2004er,Jedamzik:2005dh} that $\Xi$ could solve the primordial lithium problem.  For a choice of $\delta$ that allows $\Xi^{\pm}$ to decay around BBN, $\Xi^{\pm}$ would clearly be long-lived on detector time scales. The current bound on the production cross section of such a long -lived particle from the D0 experiment is $\lesssim 10$ fb\cite{Abazov:2008qu}.  This translates to a lower bound on the mass of a $SU(2)$-singlet $\Xi^{\pm}$ of 120 GeV.

Returning to the doublet case, we turn to Drell-Yan production of $\Xi^0 \Xi^0$.  This produces a $W^{+} W^{-} \ell^{+} \ell^{-}$ final state.  The cross section at the LHC for Drell-Yan production of the $\Xi^0$ is 93 fb at $m_{\Xi} = 200$ GeV.  In 300 fb$^{-1}$, we expect $\sim 1300$  $4 \ell + \met$ events.  While Standard Model diboson production presents an important background, lepton flavor tagging could provide a helpful discriminant: the Standard Model rate for $3 \ell \ell^{\prime}$ is very much suppressed.
It might also be possible to observe this final state in the $3\ell+2j+\met$
channel.  The Standard Model background  ($\sigma  \times$ BR) for this state from WZjj is roughly 200 fb.  So, it will be important to impose a $Z$ veto.   Fakes from top production could also be important, and more study is warranted.

Finally, $\Xi^{\pm} \Xi^0$ can be produced via an intermediate $W$ boson.  This state leads to a $Z W^{\pm} \ell^{+} \ell^{-}$ final state, which can ultimately yield striking 5$\ell$ events.  For $m_\Xi$ = 200 GeV, the production cross section is 756 fb. For $m_\Xi$ = 300 GeV, the production cross section is 176 fb, and for $m_\Xi$ = 600 GeV, the production cross section is 12 fb.   At the Tevatron  the cross section for $\Xi^{\pm}\Xi^0$ production for 
$m_{\Xi} = 175$ GeV is 76 fb, and $m_{\Xi} = 200$ GeV is 4 fb. For the lowest masses, a few multilepton events could be observed.  

\section{Models \label{sec:Models}}
One of the appealing aspects of our proposal is that it is simply incorporated into the Standard Model and almost any model of new physics.
In this section we will outline more concrete models that might realize this scenario, first in extensions of the Standard Model then in extensions to the minimal supersymmetric Standard Model (MSSM). In each case the basic building blocks are the same: namely, the new leptons states themselves, as well as an annihilation mechanism, which is accomplished with the introduction of a new scalar.
 
\subsection{Non-supersymmetric implementation}
We begin by considering a non-supersymmetric realization. The vectorlike heavy lepton mixes with the Standard Model leptons as in Eq.~(\ref{eq:mixinglagrangian}), and decays via a light lepton and $Z$ boson.  The Lagrangian is
\beq
\mathcal{L} = y_X \chi^{2} S + y_{\Xi} \bar{\Xi} \Xi S + \tilde{y} \bar{e}_{R i} \Xi h +H.c. - V(S),
\eeq
where we have rotated away a potential mass term $\ell \bar{\Xi} H$.  We also introduce a complex scalar $S= s +i a$ with potential
\beq
V(S) = m_S^2 S^{\dagger} S + \beta S^{\dagger} S^2 + \beta^{\ast} S S^{\dagger 2}+ \lambda_S (S^{\dagger} S)^2 ,
\eeq
which has a global $U(1)$ symmetry explicitly broken by $\beta$.  If we allow a vacuum expectation value for $S$ to give mass to the dark matter and $\Xi$, then $a$ is a pseudo-Goldstone boson with its mass controlled by the symmetry breaking parameter $\beta$. We have neglected a possible $S^{\dagger} S H^{\dagger} H$ mixing term. The presence of this term could affect the pseudoscalar phenomenology. While this term will be generated at the loop level, the radiatively generated size is suppressed by two powers of $\tilde{m}/\mu$ and a loop-factor -- small enough that it has no significant effects.

We consider the case where the dark matter annihilates through the pseudoscalar $a^{0}$.  This avoids the suppression by the velocity of the dark matter $v^2$ that occurs in the case of scalar mediation.   The annihilation cross section is  
\beq
\langle \sigma v \rangle_{\Xi \Xi} = \frac{y_X^2 y_{\Xi}^2}{64 \pi m_\chi^2} \sqrt{1-\frac{m_{\Xi}^2}{m_\chi^2}}\frac{1}{(1-\frac{m_a^2}{4 m_\chi^2})^2 + \frac{m_a^2 \Gamma_a^2}{16 m_\chi^4}}.
\eeq
In the limit that $m_a^2 / m_\chi^2 << 1$ and $2 m_{\Xi} \sim m_\chi$,
\beq
\langle \sigma v \rangle_{\Xi \Xi} \sim \left(5.56 \times 10^{-24} \frac{\rm{cm}^3}{\rm{s}} \right) y_X^2 y_{\Xi}^2 \left(\frac{300 \, \rm{GeV}}{m_\chi}\right)^2.
\eeq

Annihilation into pseudoscalars in this model could also be important, depending on the structure of the model.  If this occurs,  the annihilation into $a$ with subsequent decays $a \to e^+ e^-$ could provide a good fit to the data \cite{Cholis:2008vb,ArkaniHamed:2008qn}.  However, in the present realization, annihilation into light pseudoscalars is suppressed by $v^{2}$ since Majorana fermions can only annihilate into CP-odd states.

A simple modification of the above Lagrangian would realize the heavy/light annihilation scenario of the previous section.  For example, suppose S couples as $S \ell \bar{\Xi}$, and $\Xi \bar{\Xi}$ gets a mass from some other source. $S$ no longer gets a vacuum expectaion value, but can still mediate the Dark Matter annihilation.  

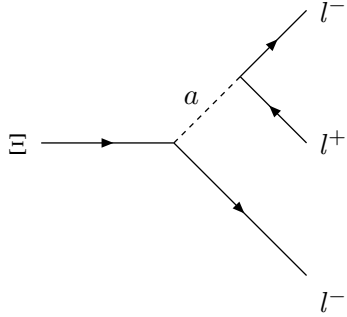
\begin{figure}
  \begin{center}
    \begin{picture}(200,200)(0,0)
      \ArrowLine(50,100)(100,100) 
      \Text(45,100)[r]{$\Xi$}
      \ArrowLine(100,100)(150,50) 
      \Text(155,45)[lt]{$\mathit{l}^-$}
      \DashLine(100,100)(125,125){2} 
      \Text(110,115)[rb]{$a$}
      \ArrowLine(150,100)(125,125) 
      \Text(155,105)[lt]{$\mathit{l}^+$}
      \ArrowLine(125,125)(150,150) 
      \Text(155,155)[lt]{$\mathit{l}^-$}
    \end{picture}
  \end{center}
  \caption{Decay of the $\Xi$ into to a lepton and a pseudoscalar $a$, which subsequently decays to leptons. \label{fig:Xidecaypseudo}}
\end{figure}

For light pseudoscalars, a new decay channel $\Xi^{\pm} \to a e^{\pm}$ opens, shown in Fig.~\ref{fig:Xidecaypseudo}.  The pseudoscalar subsequently decays to electrons.   The decays of $\Xi^{0}$ to pseudoscalars vanish in the limit of vanishing neutrino masses.  The branching ratio for charged $\Xi$ decays into light pseudoscalars is given by:
\beq
BR(\Xi \to a e) = \frac{(1 - \frac{m_a^2}{m_\Xi^2})^2}{(1 - \frac{m_a^2}{m_\Xi^2})^2 + (\frac{g_2}{c_W})^2 \frac{1}{4 y_\Xi^2}(1 - \frac{m_Z^2}{m_\Xi^2})^2 (2+ \frac{m_\Xi^2}{m_Z^2})^2}.
\eeq
This branching fraction is small if $m_a \to m_\Xi$ or if $m_\Xi >> m_Z$.  So for a low mass $\Xi$, the branching fraction to the pseudoscalar will dominate over that to $Z$ bosons until $m_a \sim m_Z$.  If this occurs, this will induce a slight modification of the results of the previous section.  The net result of $\Xi^{\pm} \to a e$ dominating is a reduction in both the necessary boost factor and the antiproton contribution. Only the $\Xi^{0} \rightarrow W e$ decays contribute to the antiprotons in this case.  This light pseudoscalar will have a very small coupling to electrons proportional to $\delta_e^2$ and thus can easily evade detection at LEP.

\subsection{A Supersymmetric Case}
The supersymmetric generalization is straightforward.
We will assume there is a dark matter sector containing the dark matter field $X$, where the fermionic component is the dark matter, that is coupled to the MSSM by a singlet $S$. This singlet then couples to a vectorlike heavy lepton.  The superpotential is given by
\beq
W = W_{MSSM} + W_{DM} + W_{\Xi},
\eeq
where the dark matter sector superpotential is
\beq
W_{DM} = y_X X X S + y_{\Xi} S \Xi \bar{\Xi}, \label{eq:DMsector}.
\eeq
In this case, the dark matter annihilates via s-channel exchange of the  the singlet field, $S$ to a vectorlike lepton.   Annihilation via  t- and u-channel processes to the singlet $S$ can also be important depending on the relatvie sizes of $y_{\Xi}$ and $y_{X}$.  


Should $\bar{\Xi}$ have the quantum numbers of a right-handed neutrinos, we have the mixing superpotential terms
\begin{equation}
W_{\Xi} = \tilde{y} \ell \bar{\Xi} H_u  + \mu \Xi {\bar \Xi}.
\end{equation}
Then the $\bar{\Xi}$ marries mostly $\Xi$ with a small amount of of the neutral component of the doublet $\ell$.  The $\Xi$ will decay via mixing with the neutrinos, so possible decays are to the $Z$ boson and a light neutrino or to a light lepton and a $W$ boson.  Decays of the $\Xi$ into a light neutrino and a $Z$ boson will produce the cosmic ray spectra of decays of $Z$ bosons.  Decays to leptons will produce a hard lepton and the cosmic ray spectra from $W$ boson decays. 
Essentially identical remarks apply to the analogous superpotential where we give $\Xi$ the quantum numbers of the right-handed electron.

We now focus on the case where the $\Xi$ has the same quantum numbers as the left-handed lepton doublet, most similar to the case discussed in the previous section.  After appropriate rotations, the leptonic portion of the superpotential can be brought to the form. 
\begin{equation}
W_{\Xi} = y \ell e_{R} H_d + \tilde{y} e \Xi H_d  + \mu \Xi {\bar \Xi}. 
\label{eq:superXi}
\end{equation}
This superpotential recovers the scenario of the previous section, assuming that the $X$ particles are the dominant component of the dark matter, and possess a $\mathbb{Z}_2$ symmetry in addition to the usual R-parity of supersymmetric models.  For the $X$ fields to dominate the dark matter density, we will need to assume that the neutralino of the MSSM sector is a subdominant component.  For example, this could be a light thermally produced wino.  

Embedding in a supersymmetric model may lead to possible new signals at the LHC. The LHC will copiously produce particles with SU(3)$_{C}$ quantum numbers, which will cascade down to the MSSM lightest supersymmetric particle.  Should cascades contain a neutralino with a large bino component, this neutralino could decay to the heavy $\Xi$ and its scalar superpartner $\tilde{\Xi}$.  This decay must compete with other decays into light fermions, however, and is likely be phase space suppressed.  Whether it is observable depends on the details of the spectrum.   Due to the approximate $\mathbb{Z}_2$ symmetry, two $\Xi$s must be produced at the end of any such cascade decay, which could give more lepton-rich signatures.  

\subsection{Realization in the NMSSM}

Alternatively, for economy, one might try to embed the supersymmetric scenario in the NMSSM, adding only a pair of  vectorlike leptons.  The $S$ field is already present to give the $\mu$ term, and the lilghtest supersymmetric particle can provide the Dark Matter.  Then we have: 
\begin{equation}
W_{NMSSM} = y \ell e_{R} H_d  + \tilde{y} e_{R} \Xi H_d  +  y_{\Xi} S \Xi {\bar \Xi} + \kappa S^{3} + \lambda S H_{u} H_{d}. \label{eq:superXiNMSSM}
\end{equation}
In this case $X$ is identified with a mixture of higgsino and singlino.  For the dark matter annihilation to be primarily into heavy leptons and not into $W$ bosons, the dark matter must have a large singlino component.  Achieving a cross section large enough to explain PAMELA requires the lightest pseudoscalar Higgs boson to be primarily singlet \footnote{It should be noted that such a large cross section will necessitate some  non-thermal mode of production of the dark matter to match the observed relic density.}. 

\section{Conclusions}
The recent evidence of an excess of high energy positrons from the PAMELA experiment raises the tantalizing possibility that we may be seeing the first hints of a dark matter signal. If so, the hard lepton spectrum combined with the absence of any antiproton excess calls out for new thinking in terms of the dark matter annihilation processes.

We have proposed a simple scenario which realizes the positron excess in PAMELA without exceeding the antiproton data.  This required the introduction of a new heavy vectorlike lepton state that mixes with the lighter Standard Model leptons.  We found that in addition to reproducing the PAMELA signal, this vectorlike lepton could be produced in Drell-Yan production at the LHC.   Even if no missing energy signals are seen at the LHC, the observation of a vectorlike lepton, when combined with astrophysical observations might still illuminate the dark matter sector.

\begin{acknowledgments}
We would like to thank Nima Arkani-Hamed, Spencer Chang, Tim Cohen, Eric Kuflik, and Jesse Thaler for useful conversations.  Thanks also to Tom Rizzo and Jesse Thaler for important comments on an earlier draft.  The work of DP is supported by the DOE under grant \#DE-FG02-95ER40899.  The work of AP is supported in part by DOE uunder grant \#DE-FG02-95ER40899 and in part by NSF CAREER grant NSF-PHY-0743315. NW is supported by NSF CAREER grant PHY-0449818 and DOE OJI grant \#DE-FG02-06ER41417. 
\end{acknowledgments}

\bibliography{vectorleptons}
\bibliographystyle{apsrev}
\end{document}